\documentclass[11pt]{article}

\usepackage[letterpaper,margin=1in]{geometry}
\usepackage[T1]{fontenc}
\usepackage{xcolor}
\usepackage{fancyvrb}
\IfFileExists{upquote.sty}{\usepackage{upquote}}{}
\usepackage[authoryear,round,longnamesfirst]{natbib}
\bibpunct{(}{)}{;}{a}{}{,}
\makeatletter
\newcommand\code{\bgroup\@makeother\_\@makeother\~\@makeother\$\@codex}
\def\@codex#1{{\normalfont\ttfamily\hyphenchar\font=-1 #1}\egroup}
\makeatother
\let\proglang=\textsf
\newcommand{\pkg}[1]{{\fontseries{m}\fontseries{b}\selectfont #1}}
\newcommand{\email}[1]{\href{mailto:#1}{\normalfont\texttt{#1}}}
\DefineVerbatimEnvironment{Code}{Verbatim}{}
\DefineVerbatimEnvironment{CodeInput}{Verbatim}{fontshape=sl}
\DefineVerbatimEnvironment{CodeOutput}{Verbatim}{}
\newenvironment{CodeChunk}{}{}
\usepackage{lmodern}
\usepackage{booktabs}
\usepackage{array}
\newcolumntype{L}[1]{>{\raggedright\arraybackslash}p{#1}}
\usepackage{graphicx}
\usepackage{tikz}
\usetikzlibrary{positioning,arrows.meta}
\shortcites{Gonthier+Asperti+Avigad:2013,Hales+Adams+Bauer:2017,Yang+Swope+Gu:2023,Xin+Guo+Shao:2024,
  Wu+Jiang+Li:2022,Azerbayev+Piotrowski+Schoelkopf:2023,Jiang+Welleck+Zhou:2023,Liang+Zhang+Cao:2024,
  Carpenter+Gelman+Hoffman:2017}

\usepackage{amssymb}
\DeclareUnicodeCharacter{211D}{\ensuremath{\mathbb{R}}}
\DeclareUnicodeCharacter{2115}{\ensuremath{\mathbb{N}}}
\DeclareUnicodeCharacter{2203}{\ensuremath{\exists}}
\DeclareUnicodeCharacter{03A9}{\ensuremath{\Omega}}
\DeclareUnicodeCharacter{03B1}{\ensuremath{\alpha}}
\DeclareUnicodeCharacter{03B2}{\ensuremath{\beta}}
\DeclareUnicodeCharacter{2082}{\ensuremath{_2}}
\DeclareUnicodeCharacter{03C4}{\ensuremath{\tau}}
\DeclareUnicodeCharacter{03C9}{\ensuremath{\omega}}
\DeclareUnicodeCharacter{2227}{\ensuremath{\wedge}}
\DeclareUnicodeCharacter{2208}{\ensuremath{\in}}
\DeclareUnicodeCharacter{2211}{\ensuremath{\sum}}
\DeclareUnicodeCharacter{2032}{\ensuremath{'}}
\DeclareUnicodeCharacter{2200}{\ensuremath{\forall}}
\DeclareUnicodeCharacter{2192}{\ensuremath{\to}}
\DeclareUnicodeCharacter{2265}{\ensuremath{\geq}}
\DeclareUnicodeCharacter{2264}{\ensuremath{\leq}}
\DeclareUnicodeCharacter{00AC}{\ensuremath{\neg}}
\DeclareUnicodeCharacter{00B7}{\ensuremath{\cdot}}

\usepackage{hyperref}
\makeatletter  
\newcommand\@doi{doi:\discretionary{}{}{}\begingroup\urlstyle{tt}\Url}
\newcommand{\doi}[1]{\href{https://doi.org/#1}{\normalfont\texttt{\@doi{#1}}}}
\makeatother
\hypersetup{colorlinks=true, linkcolor={blue!50!black}, citecolor={blue!50!black}, urlcolor={red!50!black},
  pdftitle={KAPPA: Checking the Proofs in Statistics Papers with Lean 4 and Language Model Agents}, pdfauthor={Zihua She, Xiao Wang}, pdfkeywords={proof checking, formal verification, Lean, language models, statistical theory, reproducibility}}

\title{\pkg{KAPPA}: Checking the Proofs in Statistics Papers with \proglang{Lean}~4 and Language Model Agents}
\author{Zihua She \qquad Xiao Wang\\[1ex]
  \normalsize Department of Statistics, Purdue University, West Lafayette, IN 47907, USA\\
  \normalsize \email{she8@purdue.edu}, \email{wangxiao@purdue.edu}}
\date{}

\begin{document}

\maketitle

\begin{abstract}
Papers in statistical theory rest on long proofs that few readers check line by line. We present \pkg{KAPPA} (Kit of
  Agents for Proof-checking Papers with a proof Assistant), open-source software that organizes language model agents, run by \pkg{Claude Code}, to translate the proofs of a {\LaTeX} paper into
  the proof assistant \proglang{Lean}~4 with its library \pkg{Mathlib} and check them there, result by result. Each
  numbered result receives one of four verdicts: verified unconditionally; verified conditionally, with every cited
  or unproved input stated as a named hypothesis; issue recorded, with its place in the paper and the side at fault;
  or unfinished, with the reason. The kit follows the paper's own proof step by step, rather than finding a new one. Separate agents write and
  review, each statement is compared with the paper before it is proved and fingerprinted afterwards, and every
  verdict rests on evidence that can be rerun without a language model. The kit runs unattended for hours, waits out
  usage limits, re-checks only what a revision changed and reports its cost by role, model and result. We describe
  the design, walk through a complete run, and apply the kit to four example papers with answer keys and to three
  published statistical papers, checked blind in their versions from before public corrections. In all three, the
  kit found the published error at the step that the correction names, together with further problems that we
  confirmed by hand. Given the same papers directly, the same language model working with \proglang{Lean} classified
  fewer results correctly on the longest paper and reported some statements narrower than the paper's as proved; no
  verdict of the kit claimed more than its \proglang{Lean} evidence showed.

\end{abstract}


\section{Introduction} \label{sec:intro}

A theorem in a statistics paper usually rests on a proof in an appendix, often several pages long, that combines a
few cited results with many routine steps. Readers and referees follow the main line of the argument, and errors in the routine steps can go unnoticed. Journals publish corrections when such errors come to light
\citep[three recent ones are][]{Henzi+Ziegel:2022b,Arab+Lando+Oliveira:2026,Livingstone+Betancourt+Byrne:2026}. Two of
these three leave the theorem true and repair its proof, one corrects a false statement, and in each the error sits in
a single step that a careful reader could have checked.

A specialist can formalize such a result in a proof assistant such as \proglang{Lean}~4 \citep{deMoura+Ullrich:2021}
with the library \pkg{Mathlib} \citep{mathlib:2020}, and a small trusted kernel then checks every step; formalizing
one statistics paper in this way can take weeks of expert work. A language model reads a proof in minutes, but its
answer is another informal argument that someone has to check.

\pkg{KAPPA} \citep{proofcheckkit}, a kit of agents for proof-checking papers with a proof assistant, combines the two. For each numbered result of a {\LaTeX} paper, agents translate the
statement and the paper's own proof into \proglang{Lean}, other agents review the translation, and the
\proglang{Lean} kernel checks the proof. The outcome is one of four verdicts (Section~\ref{sec:verdicts}). A result is
verified when the paper's proof compiles step by step. A cited lemma enters as a named hypothesis of the theorem. A
problem is recorded once a second agent has confirmed it, preferably with a counterexample in \proglang{Lean}. Where
the paper's argument fails, the kit reports the failing step and leaves the repair to the authors. The report also says what a person still has to check, and where. For a verified result, this is the statement, which the report prints beside its place in
the paper; for a conditional result, also the named inputs. An issue comes with the failing step, its class and its
evidence, so that the authors know where a repair has to start.

Three features of the task shape the design. A \proglang{Lean} proof can be correct for a statement that differs from
the printed one, so every statement is compared with the paper before it is proved and fingerprinted, which makes any
later change visible (Section~\ref{sec:pipeline}). Language model output differs from run to run, so every verdict is
tied to evidence that stays the same: compiled files, axiom lists and dependency walks that can be rerun offline
(Section~\ref{sec:repro}). Checking a research paper takes hours of agent work, so the kit runs unattended, survives
interruptions and re-checks only what a revision changed (Section~\ref{sec:unattended}).

The software consists of agent definitions, rules and \proglang{Bash} and \proglang{Python} scripts for
\pkg{Claude Code} \citep{ClaudeCode}, together with \proglang{Lean} commands for auditing. In spirit it is close to
tools that check the reliability of statistical computations, such as \pkg{accuracy}
\citep{Altman+Gill+McDonald:2007}, which find problems and make the evidence inspectable. As infrastructure it
resembles workflow tools such as \pkg{BatchJobs} \citep{Bischl+Lang+Mersmann:2015} and \pkg{Benchpress}
\citep{Rios+Moffa+Kuipers:2025}, which run many jobs unattended and keep their state in files.

As an example, in the paper on Hamiltonian Monte Carlo of Section~\ref{sec:errata}, a corollary
is called ``a direct consequence'' of a theorem whose conditions include a bound on the growth of the gradient. The
kit found that this bound fails for part of the family that the corollary covers. It proved in \proglang{Lean} that
the bound fails exactly for $2 < \beta < \log_2 6$ and holds for larger $\beta$, and it recorded the gap at its place
in the paper. The authors' corrigendum, posted seven years after the paper and not available to the run, describes
the same range.

Section~\ref{sec:related} compares the kit with formalization by experts, with language model provers and with
language model reviews of papers. Section~\ref{sec:verdicts} defines the verdicts and what each of them lets a reader
conclude, and Section~\ref{sec:firstrun} follows a complete run on a short paper. Section~\ref{sec:design} describes
the design, Section~\ref{sec:cases} the case studies, and Section~\ref{sec:evaluation} compares the kit with a direct
use of a language model and evaluates its cost, its fact check and its revisions. Section~\ref{sec:repro} explains
what can be reproduced and how, and Section~\ref{sec:limitations} states the limitations.

The kit is available at \url{https://github.com/ziwaa-se/kappa} under the MIT license. This paper describes
version~1.3.1, with which all runs were made. Version~1.3.4 adds fixes for problems that these runs revealed, and the
current release, 1.4.0, gives the kit its present name; earlier versions were released as the \pkg{TeX Proof-Check
Kit}. It needs \pkg{Claude Code}, \proglang{Lean}~4.19.0 and the
\pkg{Mathlib} revision pinned in the repository, \proglang{Python} 3.6 or later and a \proglang{Bash} shell. The
replication materials are described in Section~\ref{sec:repro}.


\section{Related work} \label{sec:related}

\emph{Formalization by experts.} Large results have been formalized completely, among them the four color theorem
\citep{Gonthier:2008}, the odd order theorem \citep{Gonthier+Asperti+Avigad:2013}, the Kepler conjecture
\citep{Hales+Adams+Bauer:2017} and, in \proglang{Lean}, a central theorem of condensed mathematics
\citep{Scholze:2022}. In probability, the central limit theorem has been formalized in \proglang{Isabelle/HOL}
\citep{Avigad+Hoelzl+Serafin:2017}, and \pkg{Mathlib} now contains measure-theoretic probability with conditional
expectation, independence and martingales. Projects of this kind are often organized by a \emph{blueprint}
\citep{leanblueprint}, a {\LaTeX} document whose numbered results are linked to \proglang{Lean} declarations, with a
dependency graph that shows what has been proved. The kit uses the same organization, with one group of declarations
per numbered result and the dependencies recorded. Its aim is narrower. A formalization project wants a general proof
that a library can reuse, and may change the argument to get one. The kit asks whether the paper's own argument
works, and uses whatever definitions encode the paper faithfully, even if they are too special to reuse.

\emph{Language model provers.} Systems that search for formal proofs with language models are evaluated mostly on
library theorems and on competition and textbook problems, stated in \proglang{Lean} or other proof assistants
\citep{Polu+Sutskever:2020, Zheng+Han+Polu:2022,
Yang+Swope+Gu:2023, Xin+Guo+Shao:2024}. Autoformalization translates informal statements into formal ones
\citep{Wu+Jiang+Li:2022, Azerbayev+Piotrowski+Schoelkopf:2023}, and informal proofs can guide the search for formal
ones \citep{Jiang+Welleck+Zhou:2023}. The kit uses the same methods but asks a different question. A prover asked
for \emph{some} proof of a statement may succeed with an argument unrelated to the paper's, and a failed search tells
the reader little about the paper. The kit holds the prover to the paper's route, so that a step that does not go
through becomes a finding about the paper, to be confirmed and classified. It treats the translation of the
statement, the main risk of autoformalization, as the step that most needs review: every statement is compared with
the printed text before any proving and fingerprinted afterwards. A general-purpose agent asked to check a paper with
\proglang{Lean} already reports failing steps. Section~\ref{sec:direct} shows which guarantees it still lacks.
Two recent agent systems formalize research papers: \pkg{MerLean} for papers in quantum computation
\citep{Ren+Li+Qi:2026}, and Theo for papers in theoretical computer science and discrete mathematics
\citep{SoltaniMoakhar+Gholami+Springer:2026}, which also reported a false lemma in one of its papers. Both admit
cited results as \proglang{Lean} axioms. The kit states every such input as a named hypothesis of the theorem that
uses it, so that a verdict names its conditions and the axiom check confirms that nothing else was assumed.

\emph{Language models reading papers.} Language models can comment usefully on manuscripts
\citep{Liang+Zhang+Cao:2024}, and a model asked to check a proof often finds a real error. But a reader has to redo the check to tell a real finding from a confident mistake. In the kit, a suspected error is recorded as an issue only after a second agent has confirmed it, preferably with a
counterexample that \proglang{Lean} accepts, and a verified result comes with a proof that anyone can recompile.

\emph{Reliable computation and reproducible workflows.} In statistical computing, \pkg{accuracy}
\citep{Altman+Gill+McDonald:2007} diagnoses numerical problems that can invalidate statistical analyses, and the reproducibility
literature asks that every number in a paper can be regenerated from archived code and data
\citep{Gentleman+TempleLang:2007, Peng:2011}. The kit applies this standard to proofs by leaving evidence that others
can rerun. Its driver, which runs many short sessions unattended and keeps all state in files, belongs with workflow
tools such as \pkg{BatchJobs} \citep{Bischl+Lang+Mersmann:2015}, \proglang{GNU Make} \citep{Baker:2020} and
\pkg{Benchpress} \citep{Rios+Moffa+Kuipers:2025}.

Table~\ref{tab:approaches} summarizes the comparison. A verified result of the kit rests on the \proglang{Lean}
kernel, as an expert formalization does. Its statement is reviewed by agents, the review is on record, and the report
lists the statements of the main results so that the authors can read them against the paper.

\begin{table}[t!]
\centering
\small
\begin{tabular}{L{2.3cm}L{2.6cm}L{2.6cm}L{2.6cm}L{3.0cm}}
\toprule
 & Expert formalization & Language model prover & Language model review & This kit \\
\midrule
What is checked & the theorem, by any proof & a given formal statement, by any proof & the text of the paper &
  the paper's statements, by the paper's proofs \\ \addlinespace
Who writes the formal statements & experts & people who set the problem & nobody & agents, reviewed by other agents
  before proving \\ \addlinespace
Evidence & \proglang{Lean} kernel & \proglang{Lean} kernel & an informal opinion & \proglang{Lean} kernel and
  recorded reviews \\ \addlinespace
When a proof fails & the expert repairs it & little information & a comment that needs checking & a located and
  classified issue, or the step where the check stopped \\ \addlinespace
Cited results & formalized or assumed & not applicable & read & explicit hypotheses, listed for a person \\
\bottomrule
\end{tabular}
\caption{Four ways of checking the proofs in a paper.}
\label{tab:approaches}
\end{table}


\section{Verdicts and what they mean} \label{sec:verdicts}

Every numbered result of the paper (theorem, proposition, lemma, corollary) receives exactly one of the verdicts in
Table~\ref{tab:verdicts}. The rules are written out in the kit's \code{VERDICTS.md}, and a validator checks every
recorded verdict against the evidence that must back it (Section~\ref{sec:records}).

\begin{table}[t!]
\centering
\small
\begin{tabular}{L{3.2cm}L{11cm}}
\toprule
Verdict & Requirements \\
\midrule
Verified unconditionally & The statement passed the review before proving; the whole statement is proved along
the paper's route; the gated statement is unchanged; the build is clean, only the three standard axioms of
\proglang{Lean} are used and no \code{sorry} is reachable; no external parameter or recorded debt remains; the final
audit passed. \\ \addlinespace
Verified conditionally & As above, but the proof takes explicit external parameters (a cited lemma, a fact the paper
asserts without proof, a clause of another result), each listed with the step that uses it. \\ \addlinespace
Issue recorded & A problem in this result's own statement or proof, confirmed independently: by a reviewed
\proglang{Lean} counterexample, or by a second agent's review of the reasoning. The record names the side at fault
(Table~\ref{tab:issues}). \\ \addlinespace
Unfinished & No conclusion yet, with the reason (for example, a missing library interface); not counted against
the paper. \\
\bottomrule
\end{tabular}
\caption{The four verdicts.}
\label{tab:verdicts}
\end{table}

Issues are classified by who is at fault (Table~\ref{tab:issues}), because authors need to know what to change.
An \code{ERROR} means that the printed conclusion is false. A \code{PAPER_ARGUMENT_GAP} means that the justification
given for a step fails while the conclusion may well be true. The formalization-side classes concern only the
\proglang{Lean} encoding. Steps that the paper leaves implicit and that any careful reader would fill in the same way
are recorded as \code{FAIR_IMPLICIT_STEP} and proved.

\begin{table}[t!]
\centering
\small
\begin{tabular}{L{4.2cm}L{10cm}}
\toprule
Class & Meaning \\
\midrule
\multicolumn{2}{l}{\emph{The paper}} \\
\code{ERROR} & The printed conclusion is false; a counterexample exists. \\ \addlinespace
\code{PAPER_ARGUMENT_GAP} & The justification given for a step does not hold; the conclusion may still be true. \\ \addlinespace
\code{PAPER_ASSUMPTION_GAP} & A step needs a hypothesis that the paper does not state and that cannot be derived. \\ \addlinespace
\code{EXPOSITION_GAP} & The conclusion is true, but some range or case is not covered. \\ \addlinespace
\code{DOMAIN_CLARIFICATION} & The objects are ambiguous, and the conclusion depends on the reading. \\
\midrule
\multicolumn{2}{l}{\emph{The formalization (not counted against the paper)}} \\
\code{ENCODING_ERROR} & The drafted \proglang{Lean} statement does not say what the paper says. \\ \addlinespace
\code{TRANSLATION_BLOCKED} & A needed interface is missing from the pinned library. \\
\bottomrule
\end{tabular}
\caption{Issue classes.}
\label{tab:issues}
\end{table}

Inputs that the paper cites or uses without proof are written as explicit \proglang{Lean} propositions, \code{def
Tool : Prop}, and appear as hypotheses of the verified theorem, never as axioms. Each is labeled with its
provenance: (i)~cited by the paper, (ii)~asserted without proof, (iii)~used but never stated, (iv)~a clause
of one of the paper's own results whose proof has a recorded issue, (v)~a standard fact with no interface in the
pinned library, and (p)~a clause of a predecessor result that is not proved yet, which is reconnected once it is.

The reports state what each verdict lets a reader conclude. A verified result means that the
\proglang{Lean} statement, which reviewers compared with the paper, is proved from the listed hypotheses. The
faithfulness of the translation rests on that review, and the report lets a reader repeat it. A conditional result is
as sound as its listed hypotheses, which the worksheet asks a person to check. The verdict stays conditional even after a person has checked the hypotheses and the check is recorded. An unfinished result records where the check stopped and why.


\section{A first run} \label{sec:firstrun}

We follow a two-lemma note, distributed with the kit as \code{examples/two-lemmas-demo}, through a complete run. Its
first lemma states $a^2 + b^2 \ge 2ab$. Its second states that $x^2 \ge x$ for every real $x$, with the proof ``the two
factors $x$ and $x-1$ always have the same sign''.

\subsection{Setup}

The commands below install the kit and a project for one paper. Here and below \code{\$} is the shell prompt.
\begin{CodeChunk}
\begin{CodeInput}
$ git clone https://github.com/ziwaa-se/kappa
$ kappa/install.sh mypaper-check --pinned
$ cd mypaper-check && lake exe cache get
$ scripts/pv_lean.sh build PVTools
\end{CodeInput}
\end{CodeChunk}
The third command downloads \proglang{Lean} and the compiled \pkg{Mathlib} once, and several projects can share
one installation. In a guided setup, a \pkg{Claude Code} chat runs these steps and explains them. The paper's
{\LaTeX} files, its PDF and its \code{.aux} file, which gives the exact numbers of the results, go into
\code{source/}. Unattended runs also need the project's permission file. It allows the kit's own commands and forbids
network access and edits to the paper (Section~\ref{sec:unattended}).
\begin{CodeChunk}
\begin{CodeInput}
$ cp .claude/settings.example.json .claude/settings.json
\end{CodeInput}
\end{CodeChunk}

\subsection{Running the check}

The check can run in a \pkg{Claude Code} chat, started with \code{/pv-setup} and \code{/pv-run}, or unattended from a
terminal:
\begin{CodeChunk}
\begin{CodeInput}
$ scripts/pv_drive.sh --setup source/demo.tex --auto-resume --report
\end{CodeInput}
\end{CodeChunk}
The driver starts a fresh agent session for each batch of at most three results, records everything in files, and
continues by itself after a usage limit. When it stops, the progress table shows the verdicts:
\begin{CodeChunk}
\begin{CodeOutput}
**2 numbered results: 1 verified (unconditional) · 0 verified (conditional)
  · 1 issue recorded · 0 unfinished.**
| No.     | ID  | Verdict                  | Conditions / issues
| Lemma 1 | L01 | Verified (unconditional) | ... no external parameters ...
| Lemma 2 | L02 | Issue recorded           | ... false on 0<x<1, witness
                                             x=1/2, confirmed by an
                                             independently reviewed Lean
                                             counterexample (B1).
\end{CodeOutput}
\end{CodeChunk}
Columns are shortened here. Lemma~1 is proved along the paper's route, with the two steps of the paper as two
obligations:
\begin{Code}
theorem L01_main (a b : ℝ) : a ^ 2 + b ^ 2 ≥ 2 * a * b := by
  have h1 : a ^ 2 + b ^ 2 - 2 * a * b = (a - b) ^ 2 := L01_S1 a b
  have h2 : (a - b) ^ 2 ≥ 0 := L01_S2 a b
  have h3 : a ^ 2 + b ^ 2 - 2 * a * b ≥ 0 := by rw [h1]; exact h2
  linarith
\end{Code}
For Lemma~2 the prover could not justify the ``same sign'' step and stopped at it, as the rules require. The issue
checker then encoded the printed statement and refuted it in \proglang{Lean}:
\begin{Code}
def PrintedLemma2 : Prop := ∀ x : ℝ, x ^ 2 ≥ x

theorem not_printedLemma2 : ¬ PrintedLemma2 := by
  intro h
  have h2 := h (1 / 2 : ℝ)
  norm_num at h2
\end{Code}
A second agent checked that \code{PrintedLemma2} is the printed claim, no stronger, and that $x = 1/2$ satisfies
every hypothesis of the paper. The issue was recorded after this check.

\subsection{What the run produces}

The project then holds the records listed in Table~\ref{tab:files}. The report, written for the authors, contains a
summary table, the verdict of every result with its evidence, every external input with what a person should check,
the issues with their places in the paper, and a list of what remains for a person to check. Before it is delivered,
an independent agent checks every statement in it against the records. In this run it found nine problems in the
first draft, among them a suggested fix for Lemma~2 that no review had made (Section~\ref{sec:factcheck}).

\begin{table}[t!]
\centering
\small
\begin{tabular}{L{5.4cm}L{9cm}}
\toprule
File & Contents \\
\midrule
\code{verification/RESULTS.json} & The state of every result: stage, verdict, conditions, evidence paths. \\ \addlinespace
\code{verification/audits/} & Gate, final-audit, mechanical-check and issue reports, one per review. \\ \addlinespace
\code{pv_work/*.lean} & Statement files, proofs, counterexamples; each compiles at the pinned versions. \\ \addlinespace
\code{verification/BLOCKERS.md} & Recorded issues, with places in the paper and classes. \\ \addlinespace
\code{verification/report/} & The author report, a list of files to check, and a worksheet for the external
inputs. \\
\bottomrule
\end{tabular}
\caption{Main records of a project.}
\label{tab:files}
\end{table}


\section{Design and implementation} \label{sec:design}

Table~\ref{tab:mechanisms} lists the mechanisms of the kit and what each of them guarantees about a verdict. The
subsections below describe them, and Section~\ref{sec:direct} shows what a direct use of a language model lacks
without them.

\begin{table}[t!]
\centering
\small
\begin{tabular}{L{5.2cm}L{7.2cm}l}
\toprule
Mechanism & Guarantee & Section \\
\midrule
Statement review before proving; fingerprints & The statement proved has been compared with the paper, and it cannot change unnoticed &
  \ref{sec:pipeline} \\ \addlinespace
Proof along the paper's route; failing steps kept as gaps & A verdict is about the paper's argument, not about some
  other proof & \ref{sec:pipeline} \\ \addlinespace
Inputs as named hypotheses, with provenance & Nothing is assumed silently; a conditional verdict says what remains to
  check & \ref{sec:verdicts} \\ \addlinespace
Separate writers and reviewers; issues confirmed by a second agent, with a counterexample when possible & No result is
  passed, and no problem recorded, on one agent's word & \ref{sec:roles} \\ \addlinespace
Mechanical audit and validator & The verdict agrees with its evidence: axioms, closure, hypotheses used, conditions &
  \ref{sec:mechanical}, \ref{sec:records} \\ \addlinespace
All state in files; offline replay & Every verdict can be rechecked without a language model & \ref{sec:records},
  \ref{sec:repro} \\ \addlinespace
Definitions by identifier; cascade decided by fingerprints & A revision redoes only what changed, and an issue is
  closed only by a new audit & \ref{sec:revisions} \\ \addlinespace
Independent fact check of the report & Every claim in the report is checked against the records & \ref{sec:factcheck} \\
\bottomrule
\end{tabular}
\caption{The mechanisms of the kit and what they guarantee.}
\label{tab:mechanisms}
\end{table}

\subsection{Roles and independence} \label{sec:roles}

\begin{figure}[t!]
\centering
\small
\begin{tikzpicture}[
  node distance=5mm and 6mm,
  every node/.style={execute at begin node={\hyphenpenalty=10000\exhyphenpenalty=10000}},
  box/.style={draw, rectangle, align=center, inner sep=3pt, minimum height=9mm, text width=#1},
  box/.default=2.8cm,
  rev/.style={box=#1, fill=black!12},
  rev/.default=2.8cm,
  io/.style={draw, rectangle, rounded corners=2pt, align=center, inner sep=3pt, text width=#1},
  stage/.style={font=\small\bfseries, align=center},
  arr/.style={-{Stealth[length=2mm]}},
  lab/.style={font=\footnotesize, inner sep=1.5pt}
]
\node[draw, rectangle, fill=black!5, align=center, text width=14.4cm, inner sep=4pt] (coord) at (7.55,1.25)
  {\textbf{Coordinator} (the main \pkg{Claude Code} session): assigns work, checks outputs, keeps the records, writes the reports};
\node[stage] (s1) at (1.65,0) {1.\ \code{/pv-setup}};
\node[stage] (s2) at (6.0,0) {2.\ \code{/pv-run}, every result};
\node[stage] (s3) at (13.45,0) {3.\ \code{/pv-report}};
\foreach \s in {s1,s2,s3} \draw[arr, densely dashed] (coord.south -| \s.north) -- (\s.north);
\node[box=3.1cm, below=3mm of s1] (arch) {\textbf{architect}\\ lists the numbered results, their dependencies and citations};
\node[io=3.1cm, below=12mm of arch] (in) {\textbf{Input}\\ {\LaTeX} source, PDF, \code{.aux};\\ \proglang{Lean}~4 and \pkg{Mathlib}};
\node[box, below=3mm of s2] (draft) {\textbf{drafter}\\ \proglang{Lean} statement and steps};
\node[rev, below=of draft] (gate) {\textbf{gate}\\ matches the paper?};
\node[box, below=of gate] (prove) {\textbf{prover}\\ along the paper's route};
\node[rev, below=of prove] (audit) {\textbf{auditor}\\ final audit};
\node[box=2.2cm, right=14mm of prove] (issue) {\textbf{issue checker}\\ is the paper wrong?};
\node[box, below=3mm of s3] (comp) {\textbf{compiler}\\ gathers the audits};
\node[rev, below=of comp] (fact) {\textbf{fact check}\\ checks the report against the records};
\node[io=2.8cm, below=8mm of fact] (out) {\textbf{Output}\\ author report\\ files to check\\ worksheet of inputs\\ progress table};
\draw[arr] (in) -- (arch);
\draw[arr] (arch.east |- draft) -- node[lab, above] {in order} (draft.west);
\draw[arr] (draft) -- (gate);
\draw[arr] (gate) -- node[lab, left] {pass} (prove);
\draw[arr] (gate.east) to[bend right=60] node[lab, right] {fail} (draft.east);
\draw[arr] (prove) -- node[lab, above, align=center] {step not\\ justified} (issue);
\draw[arr] (prove) -- (audit);
\draw[arr] (issue.south) |- node[lab, pos=0.3, right] {reviewed} ([yshift=2mm]audit.east);
\draw[arr] ([yshift=-2mm]audit.east) -| node[lab, pos=0.25, below] {verdicts} ([xshift=3mm]issue.east) |- (comp.west);
\draw[arr] (comp) -- (fact);
\draw[arr] (fact) -- (out);
\end{tikzpicture}
\caption{The three stages of a check and the agents of each. Shaded boxes are reviews by agents other than the writer. The architect lists the results; every result then
passes through drafter, gate, prover and auditor, with the issue checker called when a step of the paper is not
justified; the compiler and the fact check produce the report.}
\label{fig:workflow}
\end{figure}
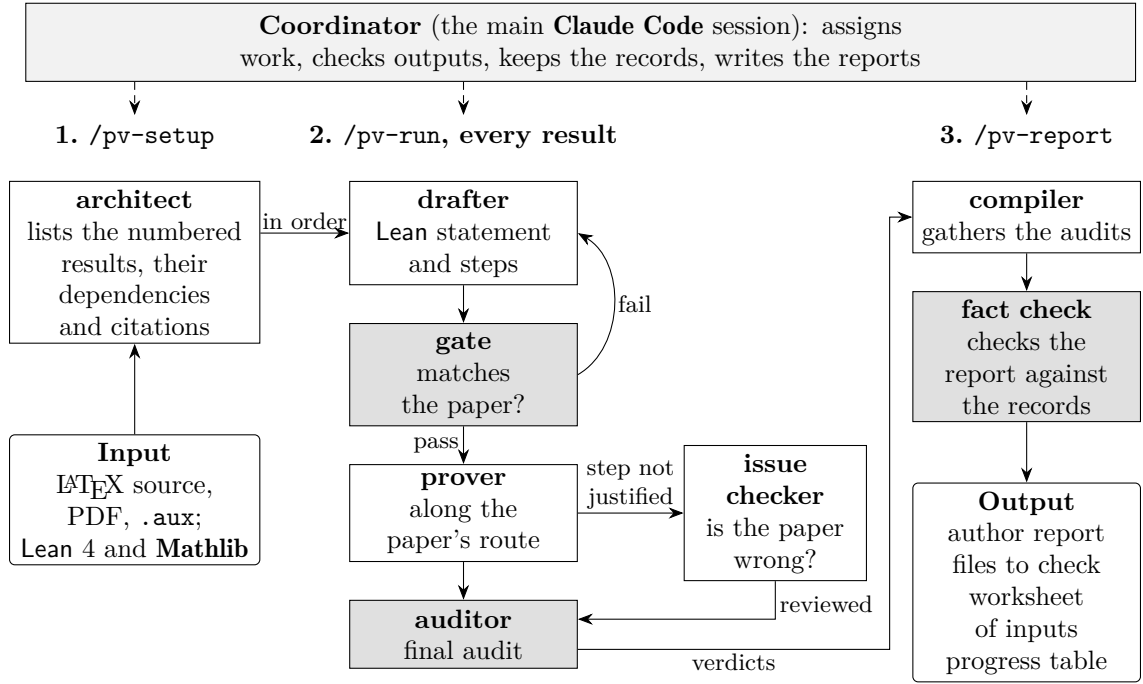

The main \pkg{Claude Code} session acts as coordinator. It assigns work, checks outputs, keeps the records and
writes the reports, and it leaves all proving and reviewing to subagents, each with its own context
(Figure~\ref{fig:workflow} and Table~\ref{tab:roles}). Writers and reviewers are always different agents, so no agent
reviews its own work. The reviewing roles run on the stronger of two models, the writing roles on the cheaper one, and a proof that gets stuck is retried on the stronger model.

\begin{table}[t!]
\centering
\begin{tabular}{lll}
\toprule
Role & Model & Writes \\
\midrule
architect & Sonnet & inventory of results, dependencies, citations \\
drafter & Sonnet & \proglang{Lean} statement and step obligations \\
gate & Opus & review of the statement, before proving \\
prover & Sonnet, then Opus & the proof, along the paper's route \\
auditor & Opus & final audit; counterexample and reasoning reviews \\
issue checker & Opus & ruling on a suspected issue; counterexamples \\
compiler & Sonnet & material for the report (read-only) \\
fact-check & Opus & check of the report against the records (read-only) \\
\bottomrule
\end{tabular}
\caption{Roles. The coordinator runs on Sonnet.}
\label{tab:roles}
\end{table}

Each role is defined in a Markdown file that carries a short rule card, the excerpts of the rules that the role
needs, each tagged with its source section, so that an agent reads only the rules it uses. A script checks
that every cited section still exists. Rule cards and the model choices together cut the cost of a run by about
three quarters without changing a verdict (Section~\ref{sec:benchmark}).

\subsection{The pipeline for one result} \label{sec:pipeline}

\begin{figure}[t!]
\centering
\resizebox{\textwidth}{!}{
\begin{tikzpicture}[
  node distance=6mm and 11mm,
  stage/.style={draw, rounded corners=2pt, align=center, font=\small, minimum height=14mm, text width=25mm, fill=black!4},
  check/.style={stage, fill=black!12},
  res/.style={draw, align=center, font=\small, minimum height=10mm, text width=27mm},
  arr/.style={-{Latex[length=2mm]}, thick},
  lab/.style={font=\footnotesize, inner sep=2pt}]
  \node[stage] (draft) {\textbf{Draft}\\drafter:\\statement, steps};
  \node[check, right=of draft] (gate) {\textbf{Gate}\\gate: vs.\ the paper;\\fingerprints};
  \node[stage, right=of gate] (prove) {\textbf{Prove}\\prover:\\the paper's route};
  \node[check, right=of prove] (audit) {\textbf{Final audit}\\auditor and\\mechanical checks};
  \node[res, right=14mm of audit] (ver) {Verified uncon-\\ditionally or\\conditionally};
  \node[stage, below=16mm of prove] (issue) {\textbf{Issue ruling}\\issue checker:\\counterexample};
  \node[check, right=of issue] (review) {\textbf{Review}\\auditor};
  \node[res, right=16mm of review] (iss) {Issue recorded};
  \node[res, below=16mm of gate] (unf) {Unfinished,\\with the reason};
  \draw[arr] (draft) -- (gate);
  \draw[arr] (gate) -- node[lab, above] {pass} (prove);
  \draw[arr] (gate.north) to[bend right=30] node[lab, above] {fail: redraft} (draft.north);
  \draw[arr] (prove) -- (audit);
  \draw[arr] (audit) -- node[lab, above] {pass} (ver);
  \draw[arr] (prove) -- node[lab, right] {suspected issue} (issue);
  \draw[arr] (issue) -- (review);
  \draw[arr] (review) -- node[lab, above] {confirmed} (iss);
  \draw[arr] (prove.south west) -- node[lab, above left] {step stopped} (unf.north east);
\end{tikzpicture}}
\caption{The stages of one result. Shaded stages are reviews by agents other than the writer; the gate records fingerprints of the statement, which the final audit compares. A result whose predecessors are not yet proved uses their clauses as parameters and is reconnected later (not shown).}
\label{fig:pipeline}
\end{figure}
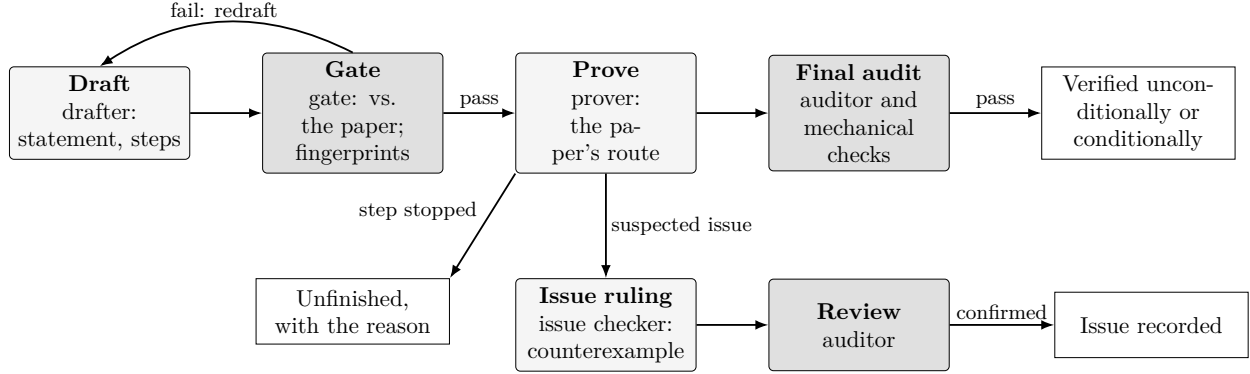

Figure~\ref{fig:pipeline} shows the stages a result passes through. The drafter writes two files: a statement file
with the definitions that the statement needs, and a draft with the main theorem and one obligation per step of the
paper's proof, each with \code{sorry} in place of a proof. The gate then compares the draft with the paper:
hypotheses, quantifiers, domains, constants, and whether every step corresponds to a sentence of the proof. A
statement that fails goes back to the drafter.

When the gate passes a statement, it records a fingerprint of every declaration: a hash of the type, a hash of the
value for definitions, and the binder annotations. The prover then proves the obligations along the paper's route.
Where the paper gives no valid justification for a step, the prover leaves that step as a gap, recorded separately, and goes on to check the remaining steps. After at most two
proof sessions per pass, the final auditor recompiles everything privately, compares the fingerprints with the
gate's, and runs the mechanical checks of Section~\ref{sec:mechanical}. A suspected problem in the paper goes to the
issue checker, whose ruling, with a counterexample when one can be built, is reviewed by an auditor before it is
recorded.

A result that depends on another result not yet proved can start as soon as the other result's statement has passed
its gate. The clause it uses enters as a predecessor parameter (p), and once the predecessor is proved, a
reconnection replaces the parameter by the predecessor's theorem. The reconnection is audited like a proof.

\subsection{Mechanical evidence} \label{sec:mechanical}

The checks that decide a verdict are deterministic. A \proglang{Lean} file of the kit, \code{PVTools.lean}, provides commands that
list the axioms a declaration depends on, walk its dependency closure, report which hypotheses its proof uses, and
print fingerprints. The auditor runs all of them with one script, \code{pv_audit_mech.py}, which writes a report of
twelve numbered items: the compile check, a private rebuild with identical output, the integrity of the audit tool
itself, the fingerprints against the gate snapshot, the axioms, the absence of \code{sorry} in the closure, the
project constants reached with the audit record that covers each, the helper declarations to review, hypothesis
usage, the conditions traced from the gated theorems, and who last wrote the audited files.

A condition listed in the records need not be used by the final proof, because a hypothesis can stay in a
signature after the step that needed it was rewritten. The script therefore walks each gated theorem and reads its hypothesis usage, and it treats an external input as a condition only when a hypothesis
that the proof uses mentions it. Inputs mentioned only by unused hypotheses are listed as formal, and inputs
registered for the result but never reached are flagged. The auditor confirms the list, and the final audit ends
with a machine-readable line such as \code{CONDITIONS: P01_NesbittInequality}, from which the records are written.

\subsection{Records and validation} \label{sec:records}

All state is in files, so that any session can stop at any time and a new one can continue. The central file,
\code{RESULTS.json}, holds every result's stage, verdict, conditions and evidence paths. A validator checks every
verdict against what must back it. A verified verdict needs a gate report and a final audit whose last lines say
\code{PASS}, a main theorem that exists in the file named, no gap, and conditions that agree with the audit's
\code{CONDITIONS} line. An issue needs a recorded problem of this result, with a class on the paper's side and a
confirming review. The coordinator writes records only through commands of \code{pv_status.py}, which fill in evidence paths from the
latest reports and write the audit log, so that records are never copied by hand.

\subsection{Revisions} \label{sec:revisions}

When the authors correct the paper, the kit compares the new version with the copy it kept and redoes only what the
change affects. A result is redone when its statement, its proof or a definition it uses changed. Definitions are
compared by identifier: every shared definition has an identifier and an exact line range, and every result lists
the definitions it uses, so that an unchanged formula next to a changed one does not affect its users.

The results that use a redone result need more care. A result consumes only the statements of the results it
imports. If a redone result passes its gate with the same statement as before, every result that imports it still consumes exactly what it consumed before, so its proof remains valid and only needs to be compiled against the new module. The kit therefore marks the direct dependents ``re-check pending'' and decides how to treat them once the redone result has passed its gate again. If the fingerprints of its definitions and main theorem are unchanged, the dependents are
rebuilt and checked mechanically by an auditor. If they changed, the dependents are redone, and their own dependents
are marked in turn. An earlier issue is resolved only by a new final audit of the revised result. The report then
shows, for each re-checked result, the verdict before and after.

\subsection{Unattended operation} \label{sec:unattended}

A long coordinator session becomes expensive, because each step re-reads the whole history. The driver
\code{pv_drive.sh} therefore runs the check in short headless sessions of at most three results each, passing all
state through the records. It stops when every result has a verdict or a recorded reason, or when a session makes
no progress, and its last line and a status file name the reason. With \code{--auto-resume} it reads the reset time
from the account's usage-limit message, sleeps until then and continues. With \code{--report} it writes the report at
the end.

Unattended agents need permissions that a person cannot grant on the spot. The project's permission file allows the kit's own scripts, in every form in which the agents call them, and refuses everything else, including network access and edits
to the paper and to the rules. A hook checks every write and every shell command against the role that makes it:
nobody writes outside the project, the coordinator writes no proof and no review, and reviewers change nothing they
review. The hook also logs each write with its role, which is how the mechanical report knows who wrote the audited
files. After each session the driver counts the permission refusals in the transcripts and stops if commands that
should have run were refused.

\subsection{Cost accounting} \label{sec:cost}

A script, \code{pv_usage.py}, reads the transcripts of all sessions and agents and sums their token use by role,
model and result, at list prices. On a subscription plan the figures measure how fast the usage limit is consumed. For the driver's sessions, the script compares its estimate with the cost that \pkg{Claude Code}
itself reports, per model. The two agree when the price table is current.


\section{Case studies} \label{sec:cases}

The case studies test whether the kit reaches the right verdicts and whether it finds errors known to exist in
published work. Four short example papers were written for this purpose, each with an answer key, and together their
keys call for every verdict and every issue class that concerns the paper; Table~\ref{tab:coverage} shows where each verdict, class
and mechanism of Section~\ref{sec:design} occurred in the runs. Three published
papers with public corrections test the kit on real proofs whose errors other people found.

\subsection{How the case studies were run} \label{sec:protocol}

\emph{Answer keys first.} For each example paper, and for each published case, we wrote down before the run what a
correct check should report for every result: the verdict, the issue class or the provenance of the inputs, the
outcomes that are also acceptable and those that are not. Where the classification is ambiguous (for
instance, a step that both needs an unstated hypothesis and asserts something false), the key accepts both classes
and says why.

\emph{Blind runs.} Each run took place in its own project folder that contained only the paper. The answer keys, the
published corrections and the repository of the kit were kept elsewhere, the project's permission file denied the
agents access to them, and network access was denied as for every run. After each run, a script
(\code{examples/tools/blind_audit.py}) scanned every transcript, of the coordinator and of all subagents, for access to
those locations, mentions of an answer file and network tools. Over 443 transcripts with 11\,341 tool calls it found
none.

\emph{Released version.} All runs used version~1.3.1 as released (commit \code{338b9d9}), with the same
\proglang{Lean} environment, the complete compiled \pkg{Mathlib}, the unattended driver with \code{--auto-resume
--report}, and the models of Section~\ref{sec:benchmark}. A pilot run of the first example on the pre-release
version is reported separately.

\emph{Scoring.} A result \emph{agrees} with the key when its verdict and classification are the expected ones or
among those the key accepts. It is \emph{found with a different class} when the kit reports the problem under another
class, and \emph{missed} or a \emph{false alarm} otherwise. We checked by hand every finding that the key did not
anticipate.

\emph{Offline confirmation.} For every run, the replay script of Section~\ref{sec:repro} rebuilt the archived
\proglang{Lean} files at the pinned versions and reran the mechanical audit, without a language model.
Every verdict whose evidence is a \proglang{Lean} file was confirmed this way, in two to thirteen minutes per run on the machine described in the Computational details. Two issues rest on a reviewed argument, and the replay has no file to rebuild
for them.

\subsection{Example papers with known answers} \label{sec:examples}

Table~\ref{tab:coverage} lists the verdicts, classes and mechanisms that the runs of this section and the next
exercised, and where. The four papers were written to exercise them, and the table records the paths that the
runs actually took.

\begin{table}[t!]
\centering
\small
\begin{tabular}{L{6.6cm}L{7.6cm}}
\toprule
Feature & Where it occurred \\
\midrule
Verified unconditionally & E1 2.1, 2.2, 2.4; E2 2.1, 2.3, 2.5; E3 2.1, 2.4; E4 2.1; B 2.5, 4.12, 5.3; C 5.15 \\ \addlinespace
Conditional on a cited result, (i) & E1 2.3; E4 2.2, 2.3; A 3.3 \\ \addlinespace
Conditional on an asserted fact, (ii) & A 3.4; C 5.14 \\ \addlinespace
Conditional on a clause of a result with an issue, (iv) & E3 2.3; B 3.3; C 5.17, 5.14, 2.2 \\ \addlinespace
Conditional on a standard fact missing from the library, (v) & C 5.14, 2.2 \\ \addlinespace
\code{ERROR} with a \proglang{Lean} counterexample & E1 2.5; E2 2.4; A 3.5; B 2.8, 4.8 \\ \addlinespace
\code{PAPER_ARGUMENT_GAP} & E1 2.6; E2 2.6; E3 2.2; A 3.1, 3.2; B 3.1, 3.2; C 5.16, 2.3 \\ \addlinespace
\code{PAPER_ASSUMPTION_GAP} & B 2.3, 4.6, 4.10, 5.2 \\ \addlinespace
\code{DOMAIN_CLARIFICATION} & E2 2.2; B 3.2 \\ \addlinespace
Fair implicit step filled in, no issue & E1 2.2, 2.4; E2 2.3; E4 2.3 \\ \addlinespace
Unfinished, \code{BUDGET_EXHAUSTED}, then a second pass & E4 2.3 \\ \addlinespace
Gate rejects an unfaithful draft & B 4.6 \\ \addlinespace
Gate requires amendments before proving & E2 2.3; B 2.5, 3.2, 4.12; A, C several \\ \addlinespace
Review rejects an issue ruling, which is then revised & B 3.2 \\ \addlinespace
Traced hypothesis usage removes a formal condition & E3 2.4 \\ \addlinespace
Revision: reset of changed results only, definitions compared by identifier, issue resolved by a new audit & E3,
  A, B, C, second round \\ \addlinespace
Reconnection of a clause after its predecessor is verified & B 3.3, C 5.17 (second round) \\ \addlinespace
Formalization-side \code{ENCODING_ERROR}, stop as \code{TRANSLATION_BLOCKED}, nothing counted against the paper & C 2.2
  (second round) \\ \addlinespace
Unattended driver waiting out a usage limit & all runs \\ \addlinespace
Independent fact check of the report & all runs \\
\midrule
Not exercised & \code{EXPOSITION_GAP} as a final verdict; (iii); a suspected issue left unconfirmed \\
\bottomrule
\end{tabular}
\caption{Where each verdict, class and mechanism occurred in the case studies (E1--E4: example papers; A--C:
published cases; first round unless stated).}
\label{tab:coverage}
\end{table}

\subsubsection{Sample variance: Errors, gaps and inherited conditions} \label{sec:e1}

The first paper collects six elementary facts about the sample mean $\bar x$ and the sample variance $s^2$ of $n$
observations. Two are identities. One bounds $s^2$ for data in an interval and rests on a cited inequality
(Popoviciu's). Samuelson's inequality $|x_i - \bar x| \le (n-1) s / \sqrt n$ is proved by dividing by $s$, without the
case $s = 0$. The lemma $s^2 \le \frac1n \sum_i x_i^2$ is false, because its proof divides by $n$ where it should
divide by $n-1$, and the last result, $s^2 \le 2$ when all $|x_i| \le 1$, is true but proved from the false lemma.
Table~\ref{tab:e1} compares the run with the answer key.

\begin{table}[t!]
\centering
\small
\begin{tabular}{lL{4.4cm}L{6.0cm}l}
\toprule
No. & Answer key & Run & Score \\
\midrule
2.1 & unconditional & unconditional & agrees \\ \addlinespace
2.2 & unconditional; the expansion is a fair implicit step & unconditional; the expansion recorded as a fair
  implicit step & agrees \\ \addlinespace
2.3 & conditional on the cited inequality, (i) & conditional on Popoviciu's inequality, (i) & agrees \\ \addlinespace
2.4 & exposition gap (case $s = 0$), or unconditional with that case filled in & unconditional; the case $s = 0$
  ruled a fair implicit step and filled in & accepted \\ \addlinespace
2.5 & \code{ERROR}, with a counterexample & \code{ERROR}; counterexample $n = 2$, $x = (1, -1)$ & agrees \\ \addlinespace
2.6 & conditional on the clause of 2.5, (iv) & \code{PAPER_ARGUMENT_GAP} in its own proof, with 2.5's issue upstream &
  differs \\
\bottomrule
\end{tabular}
\caption{The first example paper: the answer key, written before the run, and the verdicts of the run. The run took
82 minutes and cost \$14.54.}
\label{tab:e1}
\end{table}

The \proglang{Lean} file that refutes Lemma~2.5 states the printed clause, the instance and the failure of the
conclusion. Abridged, it reads
\begin{Code}
def PrintedClause : Prop :=
  ∀ {n : ℕ} (_hn : 2 ≤ n) (x : Sample n), Conclusion x

noncomputable def x0 : Sample 2 := ![1, -1]

theorem x0_var : sampleVar x0 = 2 := by
  simp [sampleVar, sampleMean, x0, Fin.sum_univ_two]; norm_num

theorem counterexample : ¬ PrintedClause := by
  intro h
  exact x0_not_conclusion (h (by norm_num) x0)
\end{Code}
where \code{Conclusion x} is the gated statement $s^2 \le \frac1n \sum_i x_i^2$ and \code{x0_not_conclusion} shows
that it fails at \code{x0}, where the right-hand side is $1$. The same file also refutes the weaker version of the
lemma that Corollary~2.6 consumes, with the extra hypothesis $|x_i| \le 1$, because $x_0$ satisfies it.

Five verdicts agree with the key or with an outcome it allows. For 2.4 the kit chose the more lenient of the two
allowed outcomes: the missing case is trivial, both sides being zero, and the kit filled it in. For 2.6 the kit went
further than the key. The proof of the corollary reads ``By Lemma~2.5, $s^2 \le \frac1n \sum_i x_i^2 \le 1 \le 2$''.
The final audit verified the corollary conditionally on the clause of 2.5, as the key expects. While compiling the
report, the compiler noticed that the audit had asked for a ruling on the intermediate claim $s^2 \le 1$ and that
none had been made. The coordinator sent the claim to the issue checker, which refuted it at the same $x_0$ and argued
that the issue belongs to the corollary itself: the corrected lemma gives $s^2 \le \frac{n}{n-1}$, so no correction of
Lemma~2.5 makes the printed chain true. We checked this by hand and agree with the kit. The key had excluded an issue
because the corollary's proof seemed fine apart from the cited lemma. That was too generous, but the score stays ``differs'' because keys are not changed after a run. In the pilot run on the pre-release version the corollary
received the key's verdict, so the classification of this borderline step varied between runs. Both reports tell the authors the same thing: the conclusion is not shown false, but the paper's proof does not establish it.

\subsubsection{Proportions and sums: Ambiguity, unproved facts and a quantifier trap} \label{sec:e2}

The second paper studies $V_n(p) = p(1-p)/n$, the variance of a sample proportion, for $p$ in ``the unit interval''
$I$, and the sample variance of a sum. Its six results raise the questions that a careful referee asks about wording. The paper never says
whether $I$ contains its endpoints. Every result holds under both readings except Lemma~2.2, $V_n(p) > 0$, which fails
at $p = 0$ and $p = 1$. Proposition~2.3, midpoint concavity of $\sqrt{V_n}$, rests on a fact that the paper calls well
known and leaves unproved, so the expected verdict is conditional on an input of provenance~(ii). Lemma~2.4, $V_n(p)
\le p/(4n)$, is false ($p = 1/2$). Proposition~2.5 asserts a constant $C$ with $V_n(p) \le C/n$ for all $n$. A draft
that lets $C$ depend on $n$ would be easy to prove and wrong, which tests the gate on the order of the quantifiers.
The proof of Proposition~2.6, $s^2(x+y) \le 2 s^2(x) + 2 s^2(y)$, starts from an equality that holds only when the
sample covariance of $x$ and $y$ vanishes, a hypothesis the paper does not state. The conclusion itself is true.

Five of the six verdicts agree with the key (Table~\ref{tab:e2}). For Lemma~2.2 the drafter chose the open
reading, because the proof itself says that ``both $p$ and $1-p$ are positive'', and noted that the lemma is false
under the closed one. The gate agreed, with the reason that the kit's ``representation choices must not manufacture
a paper error'', and asked for the ambiguity to be recorded. The open reading was proved, the
issue checker refuted the closed reading at $p = 0$, and the issue was recorded as a \code{DOMAIN_CLARIFICATION},
with the note that the authors must say which interval they mean. The false Lemma~2.4 was refuted at $n = 1$, $p
= 1/10$, and the issue record adds what the authors need in order to repair it: the bound holds on $(0,1)$ exactly
when $p \ge 3/4$. For Proposition~2.5 the gated statement puts the constant first,
\begin{Code}
def Conclusion : Prop :=
  ∃ C : ℝ, ∀ n : ℕ, 1 ≤ n → ∀ p : ℝ, V n p ≤ C / (n : ℝ)
\end{Code}
and the gate's report says why: ``uniformity in $n$ and $p$ is the content of the proposition''. The drafter got
the order right at the first attempt, so the gate had nothing to reject here. Case~B shows a rejection
(Section~\ref{sec:errata}). For
Proposition~2.6 the kit refuted the first step of the proof, $s^2(x+y) = s^2(x) + s^2(y)$, at $x = y = (0, 1)$,
recorded a \code{PAPER_ARGUMENT_GAP}, and said explicitly that the conclusion itself is not refuted.

\begin{table}[t!]
\centering
\small
\begin{tabular}{lL{4.6cm}L{5.4cm}l}
\toprule
No. & Answer key & Run & Score \\
\midrule
2.1 & unconditional & unconditional & agrees \\ \addlinespace
2.2 & \code{DOMAIN_CLARIFICATION} & \code{DOMAIN_CLARIFICATION}; open reading proved, closed reading refuted & agrees \\ \addlinespace
2.3 & conditional on the asserted concavity, (ii) & unconditional; the concavity proved from \pkg{Mathlib} & differs \\ \addlinespace
2.4 & \code{ERROR} & \code{ERROR}; counterexample $n = 1$, $p = 1/10$ & agrees \\ \addlinespace
2.5 & unconditional, constant before $n$ & unconditional, constant before $n$ & agrees \\ \addlinespace
2.6 & \code{PAPER_ASSUMPTION_GAP} or \code{PAPER_ARGUMENT_GAP} & \code{PAPER_ARGUMENT_GAP}; the first step refuted & agrees \\
\bottomrule
\end{tabular}
\caption{The second example paper. The run took 75 minutes and cost \$14.74.}
\label{tab:e2}
\end{table}

The one disagreement comes from a rule that version~1.3.1 left open. The drafter first encoded the ``well known'' concavity as an input of
provenance~(ii), as the key expects. The gate then derived the fact from \pkg{Mathlib} in a probe and recommended
proving it, and the final audit classified the sentence as a fair implicit step. Version~1.3.1 allowed both readings.
We prefer the key's, because a reader of the report should see that the paper asserts the fact without an argument,
and version~1.3.2 keeps an explicitly asserted fact as an input even when the library proves it. The verdict itself is
correct, since the concavity is true and its \proglang{Lean} proof is checked, and the report's table of boundary
cases records the decision. Only the summary table shows the result without conditions.

\subsubsection{Linear shrinkage: A revision, followed step by step} \label{sec:e3}

The third paper takes the estimator $c X$ of a mean $\theta \neq 0$ from one observation $X$ with variance $\sigma^2$,
whose risk is $R(c) = c^2 \sigma^2 + (1-c)^2 \theta^2$, and shows that the oracle factor $c^\ast = \theta^2 / (\theta^2
+ \sigma^2)$ attains the minimal risk $\rho = \theta^2 \sigma^2 / (\theta^2 + \sigma^2)$. Lemma~2.1 completes the
square. Lemma~2.2 states $\rho \le \min\{\theta^2, \sigma^2\}$, which is true, but its proof says that a harmonic mean
is at most the smaller of its two arguments, and derives the false intermediate bound $\rho \le \frac12 \min\{\theta^2,
\sigma^2\}$; with $\theta^2 = 1$ and $\sigma^2 = 3$, $\rho = 3/4 > 1/2$. Theorem~2.3 uses Lemma~2.2 for one of its
clauses and should therefore be verified only conditionally, on that clause. Corollary~2.4 cites Lemma~2.2 in its
proof without needing it, and the kit should find that the inherited condition is unused and drop it.

This paper is checked twice. After the first check we revised it as an author would: the proofs of 2.1 and 2.2 were
rewritten with their statements unchanged, and a sentence was added inside the definition of $R$, $c^\ast$ and
$\rho$, which moves every later line without changing a definition. The revision should then redo only 2.1 and 2.2,
rebuild 2.3 and 2.4 against the new modules without redoing them, close the issue of 2.2 with a reference to its new
audit, and reconnect the clause of 2.2 in 2.3, which then becomes unconditional.

The first check agreed with the key on all four results. Lemma~2.2 was recorded as a \code{PAPER_ARGUMENT_GAP}:
the issue checker refuted both false assertions of its proof in \proglang{Lean} (the harmonic mean of $1$ and $4$
is $8/5 > 1$; at $\theta = 1$, $\sigma = 2$, $\rho = 4/5 > 1/2$), proved moreover that the intermediate bound holds
only when $\theta^2 = \sigma^2$, and recorded that the lemma's conclusion is true, so that the class is not
\code{ERROR}. Theorem~2.3 was verified conditionally on the clause of 2.2 that it
uses. For Corollary~2.4 the drafter kept the clause of 2.2 that the paper cites, but the trace of hypothesis usage
showed that the proof of the corollary's statement does not reach it, and the final audit recorded it as a remark.
The corollary was verified unconditionally. The run took 71 minutes and cost \$9.85.

The revision was then handled as the kit's guide prescribes: the files in \code{source/} were replaced by the new
version and one unattended session followed the revision instructions. Its record of what changed reads, in part,
\begin{CodeChunk}
\begin{CodeOutput}
- Changes: new sentence main.tex:34 (remark in Definition 1.1; D01-D04
  UNCHANGED per defs-changed); L01 proof rewritten (43-52); L02 proof
  rewritten (58-61); T01, C01 text unchanged, only line shifts
- RESET (case b): L01, L02. Previous verdicts (L01 UNCONDITIONAL, L02
  ISSUE B1) are kept in history; B1 is marked RECHECK, not resolved.
- RE-CHECK PENDING (verdicts kept until L01/L02 pass the gate again):
  T01 (CONDITIONAL on T01_L02Consumed), C01 (UNCONDITIONAL).
\end{CodeOutput}
\end{CodeChunk}
The four shared definitions were compared by identifier and found unchanged, although every line after the new
sentence had moved, so no result was reset because of a definition. Only the two results whose proofs changed were
reset. The other two kept their verdicts, marked as waiting for a re-check. The driver then re-checked the reset
results, and the follow-up table of the new report summarizes the outcome:
\begin{CodeChunk}
\begin{CodeOutput}
| No. | Before (revision)        | Now                      | Status       |
| 2.1 | Verified unconditionally | Verified unconditionally | —            |
| 2.2 | Recorded issue           | Verified unconditionally | B1: resolved |
| 2.3 | Verified conditionally   | Verified unconditionally | —            |
| 2.4 | Verified unconditionally | Verified unconditionally | —            |
\end{CodeOutput}
\end{CodeChunk}
The issue of 2.2 was closed by its new final audit, and Theorem~2.3 lost its condition because it now imports the
verified lemma. The end state is the one the key describes, reached by a longer route. The key expected 2.3 and 2.4 to
be only rebuilt and checked mechanically, since the statements of 2.1 and 2.2 did not change. At the new gate of 2.2,
however, the statement file had lost two auxiliary definitions that only the old, wrong proof had used (the harmonic
mean and the false claim about it). Version~1.3.1 compared every definition of the statement file, saw a changed
statement and redid 2.3 and 2.4 in full; version~1.3.2 compares only what the dependents use. The revision took 23 minutes and cost \$7.40, against 71 minutes and \$9.85 for the first check.

\subsubsection{Empirical distribution functions: The limits of the library} \label{sec:e4}

The fourth paper tests the edge of the library. Lemma~2.1, that the empirical distribution function $F_n$ is
nondecreasing with $n F_n(t) \in \{0, \dots, n\}$, is elementary. Proposition~2.2 is the Glivenko--Cantelli theorem,
$\sup_t |F_n(t) - F(t)| \to 0$ almost surely, whose proof is the single sentence ``This is the Glivenko--Cantelli
theorem'': the theorem is named, but neither cited nor proved. The statement can be written with \pkg{Mathlib}'s independence and almost-sure convergence. The paper gives no
proof to check and the pinned library lacks the theorem, so the expected verdict is conditional on an input of
provenance~(v). Theorem~2.3 is the Kolmogorov--Smirnov limit law. Its statement needs the Brownian bridge,
which the pinned \pkg{Mathlib} does not provide, so the expected outcome is an unfinished check,
\code{TRANSLATION_BLOCKED}, that names what is missing: a gap of the library, which the report keeps apart from the paper's issues.

The run took a different and more ambitious route than the key expected, one the key allows
(Table~\ref{tab:e4}). For Theorem~2.3 the drafter defined a Brownian bridge by its
properties, without constructing one,
\begin{Code}
def IsBrownianBridge {Ω' : Type*} [MeasurableSpace Ω'] (Q : Measure Ω')
    (B : ℝ → Ω' → ℝ) : Prop :=
  (∀ t, Measurable (B t)) ∧
  (∀ ω, ContinuousOn (fun t => B t ω) (Icc (0:ℝ) 1)) ∧
  ∀ (m : ℕ) (u a : Fin m → ℝ), (∀ i, u i ∈ Icc (0:ℝ) 1) →
    Q.map (fun ω => ∑ i, a i * B (u i) ω) =
      ProbabilityTheory.gaussianReal 0
        (Real.toNNReal
          (∑ i, ∑ j, a i * a j * (min (u i) (u j) - u i * u j)))
\end{Code}
and stated the theorem for every such process on every probability space, with Donsker's theorem and
Kolmogorov's formula for the law of the supremum as cited inputs. The gate accepted the encoding. The first proof
pass stopped on the measurability of $D_n$, a step the paper takes for granted, and the result was recorded as unfinished, \code{BUDGET_EXHAUSTED}, the one stop reason that the example papers could not be written to trigger. The
driver's second pass, on the stronger model, proved it, and the theorem became verified conditionally on the two
cited results. The report then asks a person to check what the \proglang{Lean} evidence cannot show: that the
encoded property describes the standard Brownian bridge, and that such a process exists, since a statement
quantified over all Brownian bridges would hold vacuously if the encoded property could not be satisfied.
Proposition~2.2 was verified conditionally on the Glivenko--Cantelli theorem, which the kit labeled as cited (i)
rather than as missing from the library (v); the key accepts this, since the paper names the theorem.

\begin{table}[t!]
\centering
\small
\begin{tabular}{lL{4.8cm}L{5.6cm}l}
\toprule
No. & Answer key & Run & Score \\
\midrule
2.1 & unconditional & unconditional & agrees \\ \addlinespace
2.2 & conditional on Glivenko--Cantelli, (v) or (ii); (i) acceptable & conditional on Glivenko--Cantelli, (i) &
  accepted \\ \addlinespace
2.3 & unfinished, \code{TRANSLATION_BLOCKED}; conditional with the Brownian bridge as an explicit parameter acceptable &
  conditional on Donsker's theorem and Kolmogorov's formula, the Brownian bridge defined by its properties & accepted \\
\bottomrule
\end{tabular}
\caption{The fourth example paper. The run took 81 minutes and cost \$14.32.}
\label{tab:e4}
\end{table}

\subsection{Published results with public corrections} \label{sec:errata}

We searched for published statistical results whose proofs were later corrected in public, and chose three in
which the corrected error lies in a proof (not in a typo or a numerical experiment), the affected results are few
enough to check, and \pkg{Mathlib} covers most of what the proofs need. For each, the kit checked the version from before
the correction, blind. We then compared its report with the correction and checked the corrected version through
the revision workflow.

\paragraph{Case A: Sequential e-values for forecast comparison.} \citet{Henzi+Ziegel:2022} construct e-values for
testing whether one probability forecast dominates another, first for a single period and then sequentially for
forecasts made $h$ steps ahead. Their Section~3 has five numbered results. The last (Proposition~3.5 in the arXiv
version we checked, Proposition~2 in the journal) states that the averaged lag-$h$ e-value $e_\tau$ satisfies
$\mathbb{E}(e_\tau) \le 1$ at every almost surely finite stopping time $\tau$. The authors' correction
\citep{Henzi+Ziegel:2022b} shows by an example with $h = 2$ that this is false for $h > 1$, and proves the bound for
$e_{\tau + h - 1}$ instead. The kit checked the whole paper, version~2 on arXiv, which predates the correction.

The kit found the published error. Proposition~3.5 was recorded as an \code{ERROR}. The issue checker located the
failing step, ``$\mathbb{E}(e_\tau) = \sum_t \mathbb{E}\{\mathbb{E}(e_t \mid \mathcal{F}_t) 1(\tau = t)\} \le 1$'',
and explained it: Proposition~3.4 gives only $\mathbb{E}(e_t) \le 1$, and since $e_t$ is $\mathcal{F}_t$-measurable,
$\mathbb{E}(e_t \mid \mathcal{F}_t) \le 1$ would mean $e_t \le 1$. It then built a counterexample in \proglang{Lean}
on a probability space of two points, a fair coin: the Brier score, forecasts $p = 1/4$ and $q = 3/4$, lag $h = 2$,
both outcomes $Y_3$ and $Y_4$ equal to the coin, a betting fraction of one at time~2 only, and the stopping time that stops at~3 if $Y_3 = 0$ and at~4
if $Y_3 = 1$,
\begin{Code}
def tau0 : Bool → ℕ := fun ω => if ω = true then 4 else 3
\end{Code}
for which $\mathbb{E}(e_\tau) = \frac12 (\frac32 + 1) = \frac54$. An auditor confirmed that the instance satisfies
every printed hypothesis, including the conclusion of Proposition~3.4 that the proof relies on. The report adds that
for $h = 1$ the conclusion is believed true but not shown by the printed proof, and records the issue checker's
suggestion, marked as not checked in \proglang{Lean}: restrict the proposition to $h = 1$, and for $h \ge 2$ state the
version that the paper uses later, for the e-value stopped at $\tau + h - 1$. This is the correction the authors
published.

The run also recorded two issues that the key did not anticipate, both in proofs of true theorems and both at the
edge of a parameter range. The proof of Theorem~3.1 concludes that $E(0) \in (0,1)$, although the theorem allows
$\lambda = 1$, where $E(0) = 0$. The proof of Theorem~3.2 claims that $\lim_{\lambda \to 1} g(\lambda) = -\infty$,
which fails at $\pi_1 = 1$, an alternative the theorem allows. We checked both by hand. They are real, they leave the
theorems intact, and the authors' corrected version keeps both proofs as they were. Under the scoring rule they count as differing from the key; both are correct minor findings. Proposition~3.4 was verified
conditionally, and Theorem~3.3 conditionally on the cited theorem of Gr\"unwald and coauthors, as the key expects.
The run took 130 minutes and cost \$36.36.

We then gave the kit the authors' corrected version, version~3 on arXiv, as a revision. It reset only
Propositions~3.4 and~3.5, the two results whose text had changed, and kept the issues of Theorems~3.1 and~3.2,
whose proofs version~3 leaves as they were. The corrected Proposition~3.5, $\mathbb{E}(e_{\tau + h - 1}) \le 1$, was
verified conditionally: the case $h = 1$ without any input, and the case $h > 1$ through Proposition~3.4, on the
clause of Theorem~3.1 that it uses and on the mixture representation of the scoring function that the paper asserts.
The earlier issue of Proposition~3.5 was closed by the new final audit. The second round took 83 minutes and cost
\$24.08.

\paragraph{Case B: Stochastic dominance of convex combinations.} \citet{Arab+Lando+Oliveira:2025} introduce a class of
heavy-tailed distributions and prove, in their main result (Theorem~3.1), that a variable $X$ of this class is
stochastically smaller than every convex combination $\theta_1 X_1 + \dots + \theta_n X_n$ of independent copies. The
proof is an induction on $n$. Their corrigendum \citep{Arab+Lando+Oliveira:2026} explains that the induction step
is wrong: after a chain of correct inequalities, it uses the sign of a bracketed term in the wrong direction. The
statement is true, and the corrigendum gives a new induction step. The kit checked the whole paper, version~2 on arXiv,
which predates the corrigendum.

The kit found the published error. Its issue on Theorem~3.1 is placed at the last display of the induction step:
the first inequality there, $P(\theta_1 X_1 + \dots + \theta_n X_n > x) \ge \bar F_X(x/\theta_n) + \bar
F_X(x/(1-\theta_n))$, is false, because the term dropped from the previous line is nonpositive, so dropping it gives
an upper bound and not the claimed lower bound. A \proglang{Lean} file refutes the intermediate claim (it fails at
$x = 0$ for every law of the class, and for a Pareto law at $x = 1/2$ with equal weights $1/3$ its right-hand side is
$5/3$), and the issue was recorded as a \code{PAPER_ARGUMENT_GAP}, with the theorem's conclusion not threatened. The
issue checker added, as its own suggestion and marked as not in the paper, that the final bound follows directly from
the previous line and the defining inequality of the class. We checked this by hand: with $a = F_X(x)$, $b =
F_X(x/\theta_n)$ and $c = F_X(x/(1-\theta_n))$, the previous line minus $\bar F_X(x)$ equals $a(1 + a - b - c)$, which
is nonnegative by the defining inequality. The repair is correct, and shorter than the new induction step of the
corrigendum.

Beyond the published error, the run recorded issues on eight of the twelve registered results, against one in the
key, and we checked every one by hand (Table~\ref{tab:errata-b}). Three are the same small omission. Definition~2.2 and
Definition~4.9 require $F_X(0) = 0$, but Lemma~2.3, Lemma~4.10 and Proposition~5.2 state their characterizations
without it, so their ``if'' directions fail for $X = 0$ almost surely. The journal version adds the condition to
Lemma~2.3 and Proposition~5.2. Proposition~4.6 similarly needs $X$ to be non-degenerate, and the proof of
Proposition~3.2 takes square roots of a variable that is not assumed nonnegative. Theorem~2.8 fails for a
star-shaped function that vanishes near zero, such as $h(x) = \max(x - 1, 0)$, since $h(X)$ has an atom at zero whenever $P(X \le 1) > 0$, as for the
InvSub law $F_X(x) = x/(1+x)$.
For example, take Theorem~4.8, ``if $X$ is nonnegative super-Pareto then
$X$ is InvSub''. The kit's counterexample is $X = Z + 1$, where $Z$ has the Pareto law $F_Z(z) = 1 - 1/z$ on $[1, \infty)$: $X$ is super-Pareto,
being a convex increasing function of $Z$, and has no atom at zero, but $G(s) = 1 - F_X(1/s)$ satisfies $G(1/2) = 1 >
G(1/4) + G(1/4) = 2/3$, so it is not subadditive. The step that fails is the proof's appeal to shift invariance, which lets it assume
$h(0) = 0$. The statement presumably intends laws whose support starts at zero with no atom there, which is the
repair the run suggests. None of these issues is in the corrigendum, and none affects the main result of
the paper. Apart from Lemma~2.3 and Proposition~5.2, all of them are still in the journal version, where
Theorem~2.8 and Lemma~4.10 are numbered 2.9 and 4.11. Table~\ref{tab:errata-b} also shows
that Theorem~4.2, which the paper quotes from another paper without proof, was not registered as a result but as a cited input. The key had listed it, so the scoring rule counts it as missing.

In this run the gate also rejected a first draft. The first draft of Proposition~4.6 was failed outright, for three reasons
recorded in the gate's report: the drafted quantile function contradicted the paper's definition of the generalized
inverse; a drafted external input was false as stated, which would have made the main theorem vacuous; and one step
was false only because of how the drafter had chosen the domain, ``a formalization-side manufactured
counterexample, not a paper error''. The redrafted statement passed, and the issue finally recorded for 4.6 is the
paper's own.

\begin{table}[t!]
\centering
\small
\begin{tabular}{lL{3.7cm}L{5.2cm}L{3.2cm}}
\toprule
No. & Key & Run & Our check \\
\midrule
2.3 & verified & \code{PAPER_ASSUMPTION_GAP}: $F_X(0) = 0$ missing & correct, minor \\ \addlinespace
2.5 & verified & unconditional & --- \\ \addlinespace
2.8 & verified & \code{ERROR}: fails when $h$ vanishes near $0$ & correct, minor \\ \addlinespace
3.1 & \code{PAPER_ARGUMENT_GAP} & \code{PAPER_ARGUMENT_GAP} in the induction step & the published error \\ \addlinespace
3.2 & verified & \code{PAPER_ARGUMENT_GAP} ($\sqrt X$ for signed $X$) and \code{DOMAIN_CLARIFICATION} & correct, minor \\ \addlinespace
3.3 & conditional, (iv) & conditional on the clause of 3.1 & --- \\ \addlinespace
4.2 & conditional & not registered; a cited input & quoted without proof \\ \addlinespace
4.6 & verified & \code{PAPER_ASSUMPTION_GAP}: non-degeneracy missing & correct, minor \\ \addlinespace
4.8 & verified & \code{ERROR}: counterexample $Z + 1$ & correct; the statement needs a support starting at $0$ \\ \addlinespace
4.10 & verified & \code{PAPER_ASSUMPTION_GAP}: $F_X(0) = 0$ missing & correct, minor \\ \addlinespace
4.12 & verified & unconditional & --- \\ \addlinespace
5.2 & verified & \code{PAPER_ASSUMPTION_GAP}: $F_X(0) = 0$ missing & correct, minor \\ \addlinespace
5.3 & verified & unconditional & --- \\
\bottomrule
\end{tabular}
\caption{Case B, version 2 on arXiv. ``Key'' is the answer key written from the corrigendum before the run; ``our
check'' is the hand check of every finding that the key did not anticipate. The run took 171 minutes and cost
\$65.24.}
\label{tab:errata-b}
\end{table}

For the second round we replaced the induction step of Theorem~3.1 by the corrigendum's and changed nothing else.
The kit reset Theorem~3.1 alone, marked Proposition~3.3, which uses it, for a re-check, and left the other issues
standing, since the corrigendum does not touch them. The new induction step was found ``correct as written'', and the
theorem was verified conditionally on one clause: the ``only if'' direction of Lemma~2.3, which is true (only the
``if'' direction fails) but belongs to a lemma with a recorded issue. The earlier issue of Theorem~3.1 was closed,
and in Proposition~3.3 the clause of Theorem~3.1 was reconnected to the newly verified theorem. The second round took
54 minutes and cost \$15.24.

\paragraph{Case C: Geometric ergodicity of Hamiltonian Monte Carlo.} \citet{Livingstone+Betancourt+Byrne:2019} give
conditions under which Hamiltonian Monte Carlo is, or is not, geometrically ergodic. Their Corollary~2.3(ii) states
that for the exponential family $\mathcal{E}(\beta, \alpha)$ the method is not geometrically ergodic when $\beta < 1$
or $\beta > 2$, as a consequence of Theorem~2.2, whose light-tailed part rests on Theorem~5.14. In September 2026 the
authors added a corrigendum to the arXiv version \citep{Livingstone+Betancourt+Byrne:2026}. A condition of
Theorem~5.14 compares gradients at points whose norms differ by a factor of two, and for $\mathcal{E}(\beta,\alpha)$ it
holds only when $\beta \ge 1 + \log_2 3 \approx 2.585$, so the proof of the corollary has a gap for $2 < \beta < 1 +
\log_2 3$. With a factor of three the condition holds for every $\beta > 2$, and the corollary stands. The corrigendum
also notes that the proof of Lemma~5.16 uses that condition although the lemma does not assume it. The paper is long,
so we restricted the check to Theorem~2.2, Corollary~2.3(ii), Theorem~5.14 and Lemmas~5.15--5.17. Other results of the paper that these proofs use entered as inputs, stated by the paper and not checked. The scope includes results with no
known error; because we chose it with the correction in view, this case is a weaker test than the other two.

The kit found both points of the corrigendum. For Corollary~2.3(ii) the drafter and the gate noticed that the proof,
``a direct consequence of Theorem~2.2'', needs condition~(18) for the exponential family, and the issue checker
settled the question in \proglang{Lean} in both directions: condition~(18) fails for every member of
$\mathcal{E}(\beta,\alpha)$ with $2 < \beta < \log_2 6$ and holds for $\beta \ge \log_2 6$,
\begin{Code}
theorem osc_fails {β α : ℝ} {U : E 1 → ℝ} {gU : E 1 → E 1}
    (hU : InExpFam β α U) (hgrad : IsGradField U gU)
    (hβ : 2 < β) (hβ6 : β < Real.logb 2 6) :
    ¬ HypOscillations gU

theorem osc_holds {β α : ℝ} {U : E 1 → ℝ} {gU : E 1 → E 1}
    (hU : InExpFam β α U) (hgrad : IsGradField U gU)
    (hβ1 : 1 < β) (hβ6 : Real.logb 2 6 ≤ β) :
    HypOscillations gU
\end{Code}
Since $\log_2 6 = 1 + \log_2 3$, this is exactly the range of the corrigendum. The issue was recorded as a
\code{PAPER_ARGUMENT_GAP}, with the remark that the conclusion is not shown false on that range but is not
established by the paper either; among the recorded options for a repair is to replace the factor $2$ in~(18) by a
larger fixed constant, which is what the authors did, with $3$. For Lemma~5.16 the kit recorded that the proof uses
condition~(18), which the lemma does not assume, and added that the conclusion holds under (17) alone, which an
issue checker's probe proved in \proglang{Lean}; the corrigendum makes the same two observations. The class differs
from the key's first choice (\code{PAPER_ASSUMPTION_GAP}) but is the alternative the key accepts. Theorems~2.2 and
5.14 and Lemma~5.17 were verified conditionally, with the clause of Lemma~5.16 among their conditions, and
Lemma~5.15 unconditionally, all as the key expects. The run took 154 minutes and cost \$36.37.

The corrected version, with the corrigendum's changes carried into the text, changes the statements and proofs of all
four results around Theorem~5.14 and the proof of the corollary, so the revision redid them. Lemmas~5.15--5.17 were
all verified unconditionally: the issue of Lemma~5.16 was closed by its new audit, and Lemma~5.17 lost its condition
when the clause of Lemma~5.16 that it uses was reconnected to the verified lemma. Theorem~5.14 was verified
conditionally on the proposition on rejection probabilities that the paper uses, on standard Gaussian facts, and on
the extension to a random number of leapfrog steps that the paper asserts. Corollary~2.3(ii) was verified
conditionally on the clause of Theorem~2.2 that it uses, for a fixed number of leapfrog steps, with the earlier gap
gone because the corrected condition~(18) holds for every $\beta > 2$. Theorem~2.2 itself ended unfinished, for a
reason on the formalization side: the drafted version of Theorem~5.14 for a random number of steps carried an
antecedent that the paper does not state, which the issue checker ruled an \code{ENCODING_ERROR}, and the run stopped
that part as \code{TRANSLATION_BLOCKED} rather than redraft and re-gate the statement. The report files it on the formalization side. The second round took 88 minutes and cost \$31.28.

\paragraph{What the three cases show.} In each case the kit found the published error, placed it at the step the
correction names, and classified it as the correction implies: a false statement with a counterexample in case~A, a
gap in a true theorem in cases~B and~C. In case~C it also found the second, minor point of the corrigendum, and it
determined in \proglang{Lean} the exact range of the parameter for which the printed argument fails. Each report
ended with a repair close to the authors' own: the version stopped at $\tau + h - 1$ in case~A, a larger factor
in condition~(18) in case~C, and in case~B a repair that is simpler than the published one. The kit also reported
problems that no correction mentions, two in case~A and seven in case~B, all of which we confirmed by hand; most are
missing conditions at the edge of a definition, and two, in case~B, are theorems that are false as stated (Theorems~2.8 and~4.8). We found no false alarm
in the three cases, and no published error was missed.

\paragraph{Selection and training data.} We chose the three papers knowing their corrections, among those whose
proofs \pkg{Mathlib} can mostly express. Cases~A and~B were checked in full, case~C around the corrected results. The models may have seen some of these
corrections during training: the correction of case~A appeared in 2022 and that of case~B in the spring of 2026, both
before the training-data cutoff of June 2026 of the models we used \citep{AnthropicModels}, while that of case~C was
posted in September 2026, after it. No transcript of any of the three runs, of the
coordinator or of any agent, mentions a corrigendum or a published correction. The findings also rest on more than
the agents' word: every published error is backed by a reviewed \proglang{Lean} file, a counterexample in
case~A and a refuted step in cases~B and~C, which anyone can rebuild without a language model, and the findings that
no correction mentions go beyond what the corrections contain, although the models may have seen the journal versions,
which in case~B already repair two of them. The published run records contain the \proglang{Lean} files and the reports, and the papers themselves are available from \emph{arXiv}.


\section{Evaluation} \label{sec:evaluation}

\subsection{What a direct use of a language model does not guarantee} \label{sec:direct}

A statistician can also give a paper directly to a strong model and ask it to check the proofs in \proglang{Lean}.
We did this for the papers of Section~\ref{sec:cases}, with the same model, \code{claude-opus-5-5}, in two ways. As an
\emph{agent}, it worked in \pkg{Claude Code} without the kit, in a folder with the paper and the same \proglang{Lean}
environment, told how to compile, without network access, and asked:
\begin{quote}
\emph{Please use Lean to check the paper: for each numbered result (theorem, proposition, lemma, corollary), tell me
whether its statement and its proof are correct, prove in Lean what you can, give a counterexample in Lean when
something is false, and point out any errors in the paper.}
\end{quote}
As a \emph{chat}, it received the {\LaTeX} source in one message with the same request and no tools. The prompts and
the grading rules were fixed before the runs. Each answer was graded against the answer keys and our hand checks: a
verdict counts as backed by \proglang{Lean} only if the cited theorem compiles, uses only the standard axioms, has no
\code{sorry} and states what the paper states. The grading was done by Claude agents following these written rules,
with every judgment recorded with its reason, and checked by us. The four example papers, written to illustrate the
kit, are short enough that all three uses found the planted errors and reached essentially the same verdicts on them;
the informative comparison is on the three research papers (Table~\ref{tab:direct}), where every use also found the
published error.

\begin{table}[t!]
\centering
\setlength{\tabcolsep}{5.8pt}
\begin{tabular}{lccc}
\toprule
Paper (results) & Correctly classified & Faithful in \proglang{Lean} & Hidden or without \proglang{Lean} evidence \\
\midrule
A (5) & 5 / 5 / 5 & 3 / 3 / 0 & 0 / 2 / 3 \\
B (12) & 12 / 9 / 10 & 10 / 5 / 0 & 0 / 4 / 8 \\
C (6) & 6 / 6 / 6 & 2 / 4 / 0 & 0 / 0 / 2 \\
\midrule
Total (23) & 23 / 20 / 21 & 15 / 12 / 0 & 0 / 6 / 13 \\
\bottomrule
\end{tabular}
\caption{The kit, the direct agent and the direct chat on the three published papers (entries kit / agent / chat).
``Faithful in \proglang{Lean}'': verdicts backed, without conditions, by a compiled, \code{sorry}-free theorem stating
what the paper states. The kit's other verdicts are conditional, with every input a named hypothesis of the theorem.
``Hidden or without \proglang{Lean} evidence'': verdicts that rest on an input assumed without saying so, on a
statement narrower than the paper's presented as the paper's, or on no \proglang{Lean} file at all.}
\label{tab:direct}
\end{table}

The differences come from the mechanisms of Section~\ref{sec:design}, which the paragraphs below take in turn.

\emph{The statement is compared with the paper.} In the kit, a second agent compares every statement with the printed
one before it is proved, and the fingerprints recorded at that point stop it from changing afterwards. A direct agent
chooses its own statements, and nothing marks the difference when they drift. In case~A it reported Theorems~3.2 and~3.3 as
proved in \proglang{Lean}, but its files prove Theorem~3.2 only for $p < q$, leaving out the case $p > q$ that the
paper calls analogous, and Theorem~3.3 only for $p < q$ and $\pi_1 < 1$; its summary mentions neither restriction.
Finding them took a reading of every file.

\emph{Every input is visible.} The kit turns each cited result, each fact asserted without proof and each clause of a
result with an issue into a named hypothesis of the \proglang{Lean} theorem, and the verdict lists them for a person to
check. In none of the 23 results did a kit verdict rest on an input it did not name. The direct agent rested three verdicts on a narrower
statement or an input it did not mention (in case~A, Proposition~3.4 assumes the fact that the paper asserts at the
start of the section), and in three more gave no \proglang{Lean} evidence for steps it called correct; the chat, which
cannot compile, gave none for twelve verdicts and hid an input in one more.

\emph{Every result of the paper is checked.} The kit takes each numbered result through the same pipeline, and a
suspected problem is recorded only after a second agent confirms it. On the longest paper, case~B with twelve results, the kit classified all twelve correctly, the agent nine and the chat ten. Both missed
conditions absent from the statements of Lemma~4.10 and Proposition~4.6, and the agent also that of
Proposition~5.2; the kit's reviewers had found and confirmed all three, and so had our hand check.

\emph{The verdict and its evidence agree.} The mechanical audit ties each verdict to the theorem it names, checks its
axioms, its closure and the hypotheses its proof uses, and the validator refuses a record that disagrees with the
audits; the evidence can be rebuilt offline. A direct answer is text, and its claims about its own files can be checked only by reading those files.

The direct agent sometimes went further than the paper, proving in \proglang{Lean} inputs that the kit kept as
conditions: four of the six in case~C, and in case~A the cited theorem of Gr\"unwald and coauthors for the case it
needed. The kit keeps such inputs as conditions by design, so that a verdict on the paper rests only on the
proof that the paper contains. These mechanisms make a check with the kit more expensive than a direct use of the
model, since every statement, proof and finding passes through several agents and a mechanical audit. In return,
every statement is checked against the paper before it is proved, and every verdict can be verified afterwards from
files that rebuild without a model.

\subsection{Cost: An earlier design against version 1.3} \label{sec:benchmark}

The repository contains a benchmark paper with three results: an easy lemma, a proposition that rests on a cited
inequality, and a lemma with a planted error that is false for $x < -1$. Table~\ref{tab:bench} compares two runs of
an earlier design of the kit, which used the stronger model for every role, with two runs of version~1.3, which
introduced the per-role models, the rule cards, the audit script and the driver. All four runs used the same prompt,
the same \proglang{Lean} and \pkg{Mathlib} versions and the same machine, and all four gave the verdicts of the
answer key: the lemma verified unconditionally, the proposition conditionally on the cited inequality, and the
planted error recorded as an \code{ERROR} with a reviewed counterexample.

\begin{table}[t!]
\centering
\begin{tabular}{lrrr}
\toprule
 & Earlier design & Version 1.3 & Change \\
\midrule
Cost (list price) & \$71.15, \$65.04 & \$17.73, \$13.68 & $-77\%$ \\
Tokens processed (millions) & 73.2, 64.1 & 31.2, 21.9 & $-61\%$ \\
Time, sum of sessions (minutes) & 84, 83 & 51, 30 & $-51\%$ \\
Verdicts as in the answer key & 3/3, 3/3 & 3/3, 3/3 & \\
False alarms; unfinished & 0; 0 & 0; 0 & \\
\bottomrule
\end{tabular}
\caption{Two runs of the earlier design and two of version~1.3 on the three-result benchmark paper. Changes are
between the means.}
\label{tab:bench}
\end{table}

Part of the saving comes from cheaper models for the writing roles, whose work is reviewed afterwards. Part comes
from structure: the reviewing tasks ran on the stronger model in both designs and still became 2.4 to 2.7 times
cheaper, because of the rule cards, the audit script and shorter outputs. On research papers proving and auditing
dominate the cost, which this three-result benchmark hardly exercises.

Version~1.3.1 changed robustness rather than efficiency. Two runs on the same paper, with the models available in
October 2026 (\code{claude-sonnet-5-5} and \code{claude-opus-5-5}), again gave the three verdicts of the answer key, for
\$4.21 and \$3.78 in 14 and 16 minutes of wall-clock time. In both, the auditor ran the mechanical checks itself, and
the cost estimate from the transcripts agreed with the figure reported by \pkg{Claude Code} to the cent. Since the
models had changed as well, we also ran version~1.3 twice with the newer models. The first run cost \$5.52 and gave
the three verdicts of the key. The second ended after ten minutes with two results unfinished. The gate had found the
statement of the easy lemma faithful, but the permission system refused the way it had written its compile command,
so it failed the statement twice for want of evidence, and the proposition that depends on the lemma never started.
Version~1.3.1 allows every form of the kit's commands, tells each agent the one to use, and stops the run with an
explanation when commands in their standard form are refused (Section~\ref{sec:unattended}). With the newer models, complete runs of either version cost \$3.78 to \$5.52, but only version~1.3.1 completed both runs.

\subsection{The independent fact check of reports} \label{sec:factcheck}

Reports written by the coordinator contained errors in every run, and the fact-check role found them before
delivery (Table~\ref{tab:factcheck}). The major ones were of one kind: statements that the records did not support,
such as a suggested correction that no review had proposed, or evidence attributed to the wrong agent. In the
proportions example, for instance, the draft report advised the authors to add an argument for the ``well known''
concavity; the fact check found that no review had made that suggestion and that the compiler's own notes said so,
and the sentence was replaced by what the reviews had actually ruled. The other problems were
mostly wrong line numbers, a fair implicit step listed as a typo, or a claim that the report had already passed the
fact check before the corrections were made.

\begin{table}[t!]
\centering
\begin{tabular}{lrrrr}
\toprule
Run & Major & Moderate & Minor & Total \\
\midrule
Two-lemma example & 1 & 3 & 5 & 9 \\
Full-workflow example, first report & 1 & 7 & 6 & 14 \\
Full-workflow example, report after revision & 1 & 5 & 8 & 14 \\
Sample variance (E1) & 1 & 3 & 9 & 13 \\
Proportions and sums (E2) & 1 & 3 & 7 & 11 \\
Linear shrinkage (E3), first report & 1 & 5 & 10 & 16 \\
Linear shrinkage (E3), report after revision & 1 & 4 & 9 & 14 \\
Empirical distribution function (E4) & 0 & 2 & 8 & 10 \\
Case A, first report & 0 & 7 & 11 & 18 \\
Case B, first report & 1 & 9 & 13 & 23 \\
Case C, first report & 2 & 6 & 9 & 17 \\
Case A, report after revision & 2 & 8 & 13 & 23 \\
Case B, report after revision & 1 & 3 & 13 & 17 \\
Case C, report after revision & 1 & 12 & 12 & 25 \\
\bottomrule
\end{tabular}
\caption{Problems found by the fact check in the first draft of each report.}
\label{tab:factcheck}
\end{table}

\subsection{A revision} \label{sec:revision-eval}

In the full-workflow example, the authors' correction of the benchmark's false lemma took 17 minutes and \$5.06 to
re-check: only that lemma was redone, it became verified unconditionally, and the earlier issue was closed with a
reference to the new final audit. In the shrinkage example (Section~\ref{sec:e3}) a revision that rewrote two
proofs took 23 minutes and \$7.40, a third of the first check's time, and closed the one issue; the comparison of
definitions by identifier kept a moved but unchanged definition from resetting anything, and the cascade redid two dependents that
only needed a rebuild, which version~1.3.2 now rebuilds. The three published cases add three revisions of research papers. Each reset exactly the
results whose text the correction had changed (two in case~A, one in case~B, five in case~C), closed the published
issue only with a new final audit, kept the issues that the correction does not address, and reconnected the
clauses of the newly verified results in their dependents. They took 54 to 88 minutes and cost \$15 to \$31, between 23 and 86 percent of the cost of the first check, depending on how much the correction changed.


\section{Reproducibility} \label{sec:repro}

The agents' output differs from run to run, and a new run needs a \pkg{Claude Code} account. Everything the verdicts rest on, and every table of results in this paper, can be reproduced exactly from the archived records without a language model. The records of every run of
Sections~\ref{sec:cases} and~\ref{sec:evaluation} are part of the replication materials, with the answer keys, the
hand checks and the time and checksum of each key. For the four example papers the materials also contain the
papers, and their records are in the repository's \code{examples} folder from version~1.3.2 on; the three published
papers are available from \emph{arXiv}. The replication materials are in the folder \code{replication} of the repository, \url{https://github.com/ziwaa-se/kappa}.

\begin{enumerate}
\item \emph{Offline, exactly.} With the pinned \proglang{Lean} and \pkg{Mathlib}, every archived \proglang{Lean}
  file of every run recompiles, and the mechanical checks (axioms, \code{sorry} walk, hypothesis usage, fingerprints)
  give the archived results. A script regenerates the tables of results (verdicts against the keys, costs, times and
  fact-check counts) from the archived records. None of this needs a language model.
  Replaying one case study takes two to thirteen minutes on the machine described in the Computational details, after a one-time download of the
  compiled \pkg{Mathlib} (about 5~GB).
\item \emph{Archived, as records.} The text written by agents (proofs, reviews, reports) is archived as it was
  produced, cleaned of local paths and user names, together with a summary of every session (cost, turns, models,
  times) and the driver's log. The full transcripts remain with the authors, who ran the blind audit on them.
\item \emph{The comparison.} The prompts, the answers, the \proglang{Lean} files and the gradings of the
  comparison in Section~\ref{sec:direct} are part of the materials, together with the protocol fixed before the runs.
\item \emph{New runs, similar results.} The benchmark runner repeats a run. Verdicts should agree with the answer key, while text, cost and time vary, as the pairs of runs in Table~\ref{tab:bench} show.
\end{enumerate}

The kit's scripts have an offline test suite, run on every change on Linux and macOS, and integration tests that
compile \proglang{Lean} files.


\section{Limitations} \label{sec:limitations}

The kernel checks the proof of the \proglang{Lean} statement; whether that statement says what the paper says rests on the review by the gate and the auditor. The report lists the statements of the main results with their places in the paper, so that a person can confirm
this review. A conditional verdict is only as reliable as its listed inputs, which a person has to check. A cited result that \pkg{Mathlib} happens to contain is still kept as an input. Results that need
theory missing from the pinned \pkg{Mathlib} stop as unfinished, with the missing interface named. The agents run in
\pkg{Claude Code}, a paid service, although everything the verdicts rest on can be rechecked without it
(Section~\ref{sec:repro}). The case studies cover 42 results in seven papers, each checked once, which is too few to
estimate error rates. The permission rules and the hook guard against mistakes by agents and are not a security
boundary.


\section{Summary} \label{sec:summary}

\pkg{KAPPA} turns the proofs of a {\LaTeX} paper into a \proglang{Lean} check of that paper's own
argument and reports one of four verdicts for every numbered result. Its design responds to the facts that a proof that compiles may prove the wrong statement, that language model output varies and needs checking, and that checking a paper takes hours. The kit therefore reviews and fingerprints every statement before it is proved, separates the agents
that write from those that review, rests each verdict on files that can be rebuilt without a model, and runs
unattended in short sessions that keep all state on disk.

On four example papers with answer keys written in advance, 17 of 19 verdicts agreed with the key or with an outcome
the key allowed. Of the other two, one was a finding that our hand check showed to be better than the key, and one
revealed an ambiguity in the kit's rules that the current release resolves. On three published papers checked before their
corrections, the kit found each published error at the step the correction names, with a \proglang{Lean}
counterexample or a refuted step, and suggested repairs close to the authors'. It also reported problems that the
corrections do not mention, which we checked by hand and found correct. Most are conditions missing at the edge of a definition. The same model, given the papers directly as a single agent with \proglang{Lean}, also found the
published errors, but it missed problems on the longest paper, reported statements narrower than the paper's as
proved, and assumed inputs without saying so. The mechanisms of the kit are designed to prevent each of these (Section~\ref{sec:direct}). A first check
with the kit cost between \$10 and \$65 at list prices and took one to three hours of unattended running, and a revision cost between 23 and 86 percent of that.

A verified result of the kit is backed by a \proglang{Lean} proof of a statement that was reviewed against the printed text before it was proved, and a conditional result names every input that a person still has to check. Within the reach of the
pinned \pkg{Mathlib}, the kit gives authors and referees a list of which steps of a proof hold and which need repair, each located in the paper, classified and backed by files that can be rechecked.


\section*{Computational details}

The runs in this paper used \pkg{KAPPA}~1.3.1 (commit \code{338b9d9}, released as the \pkg{TeX Proof-Check
Kit}), \pkg{Claude Code}~2.1.296 with the models \code{claude-sonnet-5-5} and \code{claude-opus-5-5}, \proglang{Lean}~4.19.0
and \pkg{Mathlib} revision \code{c44e0c8e}, on a Mac mini with an Apple M4 processor and 16~GB of memory running
macOS~26.6.2. Up to seven runs shared the machine at a time. Costs are those that \pkg{Claude Code} reported for each session, at list prices. The estimates that \code{pv_usage.py} computes from the transcripts agree with them to
within five cents per run. Times are active running times, without the waits for usage limits. The comparison of Section~\ref{sec:direct} used
\code{claude-opus-5-5} through the same \pkg{Claude Code} version. The benchmark runs of the earlier design (version~1.2.1 in the kit's changelog) and of version~1.3 used the
models \code{claude-sonnet-5} and \code{claude-opus-5}; their costs are list-price estimates computed from the
transcripts.

\section*{Acknowledgments}

Xiao Wang acknowledges support from the National Science Foundation under Grant No.~SES-2316428 and from Anthropic
through its AI for Science program.

Xiao Wang conceived the project and supervised it. Zihua She designed and implemented the software, wrote the
example papers and their answer keys, ran the case studies, checked the findings by hand and wrote the first draft
of the manuscript. Both authors chose the published cases and revised the manuscript.

\pkg{Claude Code} helped develop the kit and revise this manuscript; the authors checked all of it and are
responsible for the content.


\bibliography{refs}


\newpage

\begin{appendix}

\section{Command reference} \label{app:commands}

Most users never type these commands: in the guided mode, a \pkg{Claude Code} chat started in the kit's folder
runs them on requests such as ``I have a new paper; call it \code{mypaper}'' or ``Continue \code{mypaper}''. They are
listed here because they are the interface that the agents use and that a reader of the records may want to rerun.
Table~\ref{tab:skills} lists the chat commands, which are \pkg{Claude Code} skills, and Table~\ref{tab:scripts} the
scripts; every script prints its options when called without arguments.

\begin{table}[!ht]
\centering
\small
\begin{tabular}{lL{11.5cm}}
\toprule
Command & What it does \\
\midrule
\code{/pv-setup} & checks the \proglang{Lean} environment, has the architect list every numbered result with its
  place in the paper, its dependencies and its citations, and writes the records \\ \addlinespace
\code{/pv-run} & the coordinator's loop: takes each result through draft, gate, proof and final audit, sends
  suspected problems to the issue checker, and records verdicts, until every result has one \\ \addlinespace
\code{/pv-resume} & continues after an interruption: reconciles half-finished agent outputs with the records and
  resumes the loop \\ \addlinespace
\code{/pv-revise} & handles a new version of the paper: compares it with the kept copy, resets only the affected
  results and marks their dependents for re-checking \\ \addlinespace
\code{/pv-report} & writes the author report, the worksheet of external inputs and the index of files to check, and
  has them fact-checked \\ \addlinespace
\code{/pv-status} & prints the progress table \\
\bottomrule
\end{tabular}
\caption{Chat commands.}
\label{tab:skills}
\end{table}

\begin{table}[!ht]
\centering
\small
\begin{tabular}{lL{8.8cm}}
\toprule
Script & What it does \\
\midrule
\code{install.sh} & creates a project, or upgrades one in place, keeping its records \\ \addlinespace
\code{scripts/pv_drive.sh} & the unattended driver: short headless sessions, \code{--auto-resume}, \code{--report} \\ \addlinespace
\code{scripts/pv_lean.sh} & compiles one file (\code{check}) or a module (\code{build}) with the pinned versions;
  refuses files with input-output actions \\ \addlinespace
\code{scripts/pv_status.py} & all writes to the records (\code{record}, \code{add-tool}, \code{reset},
  \code{human-check}, \code{set-deps}), the validator (\code{validate}), what can start now (\code{next}), what a
  revision changed (\code{defs-changed}, \code{followup}), and the progress table (\code{table}) \\ \addlinespace
\code{scripts/pv_audit_mech.py} & the mechanical part of a final audit (Section~\ref{sec:mechanical}) \\ \addlinespace
\code{scripts/pv_snapshot.py} & saves and restores checkpoints of the records, and keeps the copy of the paper that revisions are compared with \\ \addlinespace
\code{scripts/pv_usage.py} & token use and cost by role, model and result \\ \addlinespace
\code{scripts/pv_guard.py} & the hook that checks writes and shell commands against the role making them \\ \addlinespace
\code{scripts/pv_settings.py} & writes the permission file for the project's paths and checks an existing one \\ \addlinespace
\code{scripts/pv_cards_check.py} & checks that every rule cited in an agent's card still exists \\ \addlinespace
\code{examples/tools/replay.py} & rebuilds an archived run offline and reruns its mechanical audits \\ \addlinespace
\code{examples/tools/score.py} & compares a run with an answer key \\ \addlinespace
\code{examples/tools/blind_audit.py} & scans a run's transcripts for forbidden reads and network access \\ \addlinespace
\code{examples/tools/sanitize.py} & copies a run for publication without local paths, user names or e-mail
  addresses \\
\bottomrule
\end{tabular}
\caption{Scripts.}
\label{tab:scripts}
\end{table}

\end{appendix}

\end{document}